\pdfoutput=1
\documentclass[iop,apj]{emulateapj}
\usepackage{graphicx,epsfig}
\usepackage{rotate}
\usepackage{amsmath, amsthm, amssymb}
\usepackage{times}
\usepackage{soul}
\usepackage[plainpages=false, colorlinks=true, anchorcolor=blue, linkcolor=blue, citecolor=blue, bookmarks=false]
{hyperref}

\newcommand{\bq}{\begin{equation}}
\newcommand{\eq}{\end{equation}}

\def\gtsim{\lower.5ex\hbox{$\buildrel > \over\sim$}}
\def\ltsim{\lower.5ex\hbox{$\buildrel < \over\sim$}}

\def\sun{\mbox{$_\odot$}}
\def\apjl{ApJL}
\def\apj{ApJ}
\def\apjs{ApJS}
\def\mnras{MNRAS}

\def\aj{AJ}
\def\aap{A\&A}
\def\aaps{A\&A Suppl.}

\newcommand{\subdate}{2014 May 17}
\newcommand{\shortauth}{Chatzopoulos, Graziani, \& Couch}
\newcommand{\slugcom}{Submitted to ApJ on \subdate}
\slugcomment{\slugcom}

\lefthead{\sc \footnotesize \slugcom \hfill \shortauth}
\righthead{\sc \footnotesize \slugcom \hfill \shortauth}

\begin{document}
\title
{CHARACTERIZING THE CONVECTIVE VELOCITY FIELDS IN MASSIVE STARS}
\author{Emmanouil Chatzopoulos,\altaffilmark{1,2} Carlo Graziani,\altaffilmark{1} and Sean M. Couch\altaffilmark{1,3}}
%%%  author names
\email{manolis@astro.as.utexas.edu}
\altaffiltext{1}{Department of Astronomy \& Astrophysics, Flash Center for Computational
Science, University of Chicago, Chicago, IL, 60637, USA.}
\altaffiltext{2}{Enrico Fermi Fellow}
\altaffiltext{3}{Hubble Fellow}

\begin{abstract}

We apply the mathematical formalism of vector spherical harmonics decomposition to convective
stellar velocity fields from multi-dimensional hydrodynamics simulations, and show that the
resulting power spectra furnish a robust and stable statistical description of stellar convective turbulence. 
Analysis of the power spectra help identify key physical parameters of the convective process
such as the dominant scale of the turbulent motions that influence the structure of massive evolved pre-supernova
stars. We introduce the numerical method that can be used to calculate vector spherical harmonics power spectra
from 2D and 3D convective shell simulation data. Using this method we study the properties of oxygen shell burning
and convection for a 15-$M_{\odot}$ star simulated by the
hydrodynamics code FLASH in 2D and 3D.
We discuss the importance of realistic initial conditions to achieving successful core-collapse supernova
explosions in multi-dimensional simulations. We show that the calculated power spectra can be used
to generate realizations of the velocity fields of pre-supernova convective shells.
We find that the slope of the solenoidal mode power spectrum remains
mostly constant throughout the evolution of convection in the oxygen
shell in both 2D and 3D simulations.
We also find that the characteristic radial scales of the convective elements
are smaller in 3D than in 2D
while the angular scales are larger in 3D. 

\end{abstract}


\vskip 0.57 in

\section{INTRODUCTION}\label{intro}

Many astrophysical systems are characterized by highly anisotropic, turbulent
or even chaotic motions of their constituents. The case of  
energy transport in stellar interiors via convection is a classic example. Depending
on a star's mass and age convective instability can be triggered by nuclear burning and
may arise in several parts of the stellar interior (core or shell convection). 

Due to computational limitations the effects of convection on
stellar evolution
have predominantly been studied via the use of one-dimensional spherically symmetric stellar
evolution codes such as KEPLER \citep{1978ApJ...225.1021W, 2002RvMP...74.1015W}, 
TYCHO \citep{2005ApJ...618..908Y} the
GENEVA code \citep{2008Ap&SS.316...43E} and more recently the
Modules for Experiments in Stellar Astrophysics (MESA)
\citep{2011ApJS..192....3P, 2013ApJS..208....4P}. 
All of these codes use the standard mixing-length theory (MLT) to treat
convection based on either the Schwarzschild or the Ledoux criteria plus
parametrized treatments for the effects of semi-convection, convective overshoot
and thermohaline mixing (see, e.g., \citet{2000ApJ...528..368H}). A topic of current
debate is how accurate those parametrized MLT prescriptions are
as compared to intrinsically three-dimensional simulations of convection.

Significant efforts have been made to simulate stellar convection in multiple dimensions 
over a timescale short compared to the evolutionary timescale, and to compare the output of such simulations
to the parametrized predictions. \citep{2007ApJ...667..448M} presented multi-dimensional 
simulations of oxygen shell burning and hydrogen core burning for a 23-$M_{\odot}$ 
core-collapse supernova (CCSN) progenitor star. They found significant differences between
the 2D and the 3D treatment and underscored the fact that the convective mixing regions are better
predicted using dynamic boundary conditions (BCs) rather than local and static MLT criteria. This dynamical
behavior of convective boundaries is found to be a source of gravity waves that can, under certain
circumstances, lead to episodic mass-loss in the years preceding the supernova (SN) explosion
\citep{2012MNRAS.423L..92Q, 2014ApJ...780...96S}. Such pre-SN mass loss events can give rise
to supernova impostors, as is the case with the pre-explosion outbursts of SN~2009ip 
\citep{2013MNRAS.430.1801M, 2014ApJ...780...21M, 2014ApJ...785...82S}.
Similar studies have been done in the case of main-sequence (MS) core convection \citep{2013ApJ...773..137G} and
vigorous pre-SN convection (with emphasis on energetic Si-shell burning)
in the hours prior to core collapse \citep{2006PhDT........20M,2011ApJ...741...33A},
also Couch et al. (2014, in preparation).

These multi-dimensional studies exhibit pronounced shell asymmetries and dynamical
interactions between adjacent convective regions. Such effects can significantly change the structure
not only of the pre-SN star but of its circumstellar (CS) environment \citep{2014arXiv1401.4893M},
and, as a result, can affect the initial conditions for the core-collapse process and subsequent explosion.
Indeed, it has recently been shown that the outcome of CCSN simulations can be {\it qualitatively} different
for realistic aspherical initial conditions.
\citep{2013ApJ...778L...7C} show that imprinting physically-motivated velocity
fluctuations in the convective regions of the progenitor star prior to
collapse can result in 
shock revival in 3D CCSN simulations that fail to explode otherwise.

The importance of initializing such velocity
perturbations in order to characterize the multi-dimensional nature of pre-SN convection has
been pointed out in the past and several formalisms have been proposed, 
including scalar spherical harmonics and Fourier decomposition \citep{2012arXiv1204.4842C} that
are commensurate with the Kolmogorov energy spectrum expected for highly turbulent stellar regions.
Fourier modes, however, are poor matches to the spherical boundary conditions
relevant to this problem. In addition, other approaches such as
the spherical Fourier-Bessel decomposition have attempted to analyze
scalar fields in CCSN convection \citet{2014MNRAS.440.2763F} in
spherical, concentric shells.
In this paper we introduce a mathematical framework for analyzing stochastic stellar
velocity fields, the method of decomposition into vector spherical harmonics (VSH), and apply the method to the
case of CCSN progenitor convection. VSH decomposition of multi-dimensional simulation
data can be used to extract power spectra that describe the distribution of convective power over the length
scales of the system. 
VSH power spectra can also be used to produce realizations of velocity fields that
capture the non-radial perturbations of the flow due to convection, and therefore provide more realistic initial conditions
for multi-dimensional CCSN simulations.
The toolset of VSH has been used in other fields of astrophysics where random velocity distributions are
present such as the local stellar velocity field \citep{2007AJ....134..367M} and stellar pulsations and oscillations
\citep{1969A&A.....2..390S, 1967PhFl...10.1186K}. 

Our paper is organized as follows: in Section \ref{vsh} we present the basic mathematical formulation of VSH, in Section \ref{sims}
we discuss the numerical evaluation of VSH and relevant consistency tests that illustrate
the accuracy of our results and apply the method to the case of a 2D oxygen
shell burning simulation. In Section \ref{sims_3D} we
apply VSH decomposition to a 3D oxygen shell burning simulation and discuss the differences
between the corresponding power spectra. Finally, in Section \ref{disc_2} we summarize our conclusions and
discuss the importance of this technique to setting realistic initial conditions for the CCSN simulations.

\section{THE FORMULATION OF VSH}\label{vsh}

 Much of the VSH formulation described below is based on material in 
\citet{1953mtp..book.....M}, \citet{1961hhs..book.....C}, \citet{1975clel.book.....J}
and \citet{1995mmp..book.....A}. 
Details on the derivation of the VSH modes, their
orthonormality relations, proper treatment of boundary conditions as well as dependence on data
dimensionality and simulation domain can be found in Appendix \ref{sec:appendix}. 

We seek to characterize velocity fields that we may choose to 
impose as an initial condition for a stellar simulation in a 
spherical shell $\Sigma$, consisting of the region 
$R_{1}<|\mathbf{x}|<R_{2}$ where $\mathbf{x}$ is the position vector. 
Such a field ought to satisfy certain 
physically-motivated mathematical requirements. One such requirement 
is that the framework for specifying the velocity field should allow 
good control of the divergence of the momentum density field. 
Consider the continuity equation:
\begin{equation}
\frac{\partial\rho}{\partial t}+\nabla\cdot\left(\rho\mathbf{u}\right)=0,\label{eq:MassCons}
\end{equation}
where $\mathbf{m} \equiv \rho \mathbf{u}$ is the momentum density.
If one is to specify a velocity field on top of some nearly-hydrostatic
mass configuration $\rho(\mathbf{x})$, it would be well to control
the size of $|\partial\rho/\partial t|$ so that the velocity field
does not inadvertently create large departures from the near-equilibrium
initial state. According to Equation~(\ref{eq:MassCons}), this can be
accomplished by ensuring that the divergence of the momentum density,
$\nabla\cdot(\rho\mathbf{u})$ has a controllable magnitude, which
we may set to zero (the ``anelastic'' case) or to a ``small''
value as suits the case.

A second requirement is that the velocity field imposed on the problem
should add no net momentum to the mass configuration. That is:
\begin{equation}
\int_{\Sigma}d^{3}\mathbf{x}\,\rho(\mathbf{x})\mathbf{u}(\mathbf{x})=0,\label{eq:NoNetMomentum}
\end{equation}
If this requirement were not
satisfied the result would be an initial condition that imparts unwanted kicks to the mass configuration.

These first two requirements are expressed in terms of $\mathbf{m}$, 
rather than directly in terms of
the velocity $\mathbf{u}$. This suggests that we model $\mathbf{m}$
using the set of modes described below, and obtain $\mathbf{u}$ indirectly
by $\mathbf{u}=\mathbf{m}/\rho$. This, then, is what we shall do.

A third necessary requirement relates to boundary conditions. We
demand that the radial velocity should go continuously
to zero at specified radii. In particular, when decomposing a momentum
field $\mathbf{m}(\mathbf{x})$ in a spherical shell $R_{1}<|\mathbf{x}|<R_{2}$,
we will require that $\mathbf{m}(\mathbf{x})$ should be purely tangential
at $|\mathbf{x}|=R_{1}$ and at $|\mathbf{x}|=R_{2}$. The boundary
conditions we require are thus 
\begin{equation}
\mathbf{x}\cdot\mathbf{m}(\mathbf{x})=0\textnormal{ for }|\mathbf{x}|=R_{1}\textnormal{ and for }|\mathbf{x}|=R_{2}.\label{eq:MomBC}
\end{equation}
Such a velocity field with support confined to a spherical shell
is of interest, for example, in the case of a CCSN
progenitor, with a nearly quiescent iron core surrounded by a convectively
burning silicon/oxygen shell. 

A final requirement is that the set of modes
used to express the velocity field should be orthonormal and complete.
This requirement allows us to analyze an existing velocity field in
terms of a unique spectrum in a meaningful way, and to use a spectrum
of modes to generate velocity field realizations unambiguously.

The challenge is therefore to decompose a velocity field that
satisfies the above requirements into three vector fields (one
irrotational and two solenoidal component fields). To do
so we first recall that the solutions of a self-adjoint partial differential equation (PDE) 
form a complete, orthonormal set of functions. 
One of the simplest and well-studied such PDEs is the scalar Helmholtz equation, 
$(\nabla^{2}+k^{2})\phi_{k}=0$,
wherein the solution 
$\phi_{k}(\mathbf{x})$ 
may be regarded as an
eigenvector of the Laplacian operator $\nabla^{2}$ with eigenvalue
$-k^{2}$. It is a standard result of Sturm-Liouville theory that
eigenfunctions of self-adjoint operators such as the Laplacian satisfying
specified boundary conditions may be pressed into service as complete
sets of basis functions for the expansion of quite general functions
and distributions satisfying the same boundary conditions.

The same principle is applied to the expansion of \emph{vector}
fields. That is to say, we seek our basis of vector functions
among the solutions of the \emph{vector Helmholtz equation}
\begin{equation}
(\nabla^{2}+k^{2}) {\bf Z}_{k}(\bf x) = 0.\label{eq:vec_helm_eq}
\end{equation}
These solutions can be constructed using families of solutions of
the scalar Helmholtz equation with the required boundary conditions,
which will confer their orthonormality/completeness properties upon
the vector solutions.

We note in passing that we do not ascribe any dynamical significance to
Equation (\ref{eq:vec_helm_eq}).  Rather, we are relying on the vector Helmholtz equation
merely to generate modes with useful boundary conditions for
the purpose of analyzing and realizing velocity fields at fixed times.
This is analogous to using plane wave Fourier modes --- complete,
orthonormal solutions of the Helmholtz equation in Cartesian coordinates
--- to characterize the instantaneous state of a fluid in a Cartesian
box.

A general vector field has three degrees of freedom at every
point. These could be characterized by three functions, one for each
coordinate component of the field. While such a decomposition is certainly
simple, it does not allow us to address the requirements stated above. Instead, we consider
the so-called Helmholtz decomposition, which states that any vector
field $\mathbf{Z}(\mathbf{x})$ may be decomposed into a sum of an
irrotational field $\mathbf{I}(\mathbf{x})$ and a solenoidal field
$\mathbf{S}(\mathbf{x})$, $\mathbf{Z}(\mathbf{x})=\mathbf{I}(\mathbf{x})+\mathbf{S}(\mathbf{x})$,
where $\mathbf{I}(\mathbf{x})=\nabla\Phi(\mathbf{x})$ and $\mathbf{S}=\nabla\times\mathbf{V}(\mathbf{x})$
for some scalar function $\Phi$ and vector function $\mathbf{V}$,
so that $\nabla\times\mathbf{I}=0$ and $\nabla\cdot\mathbf{S}=0$.
In this decomposition, all the divergence of the field $\mathbf{Z}$
is associated with the irrotational component $\mathbf{I}$, while
all the curl attaches to the solenoidal component $\mathbf{S}$. By
providing a set of modes for $\mathbf{I}$ separate from the modes
supplied to represent $\mathbf{S}$, the
divergence of the vector field $\mathbf{Z}$ can be separated out
and ascribed any magnitude, including zero.

Starting from an arbitrary
solution $\phi_{k}(\mathbf{x})$ of the scalar Helmholtz equation
\begin{equation}
(\nabla^{2}+k^{2}) \phi_{k} = 0.\label{eq:ScalarHelm}
\end{equation}
 we construct vector functions
\begin{eqnarray}
\mathbf{A} & \equiv & \nabla\phi_{k}\label{eq:IrrotFun}\\
\mathbf{B} & \equiv & \nabla\times\left(\mathbf{x}\phi_{k}\right)\nonumber \\
 & = & \nabla\phi_{k}\times\mathbf{x}\label{eq:SolenoidFun1}\\
\mathbf{C} & \equiv & \nabla\times\nabla\times\left(\mathbf{x}\phi_{k}\right).\label{eq:SolenoidFun2}
\end{eqnarray}

The detailed choice of $\phi_{k}(\mathbf{x})$ is deferred to Appendix \ref{sec:appendix}.
Obviously, we have $\nabla\times\mathbf{A}=0$ and $\mathbf{\nabla}\cdot\mathbf{B}=\nabla\cdot\mathbf{C}=0$,
so $\mathbf{A}$ is irrotational and $\mathbf{B}$ and $\mathbf{C}$
are solenoidal. It is straightforward to verify that these fields
are solutions of the vector Helmholtz equation.

In Appendix \ref{sec:appendix} we show that the modes satisfying the requirements stated above are given by the following 
set of equations:
\begin{eqnarray}
\mathbf{A}_{lnm}(\mathbf{x}) & = &
k_{ln}^{(N)\,-1}\Bigg[\frac{dg_{l}^{(N)}\left(k_{ln}^{(N)}r\right)}{dr}\frac{\mathbf{x}}{r}Y_{lm}(\mathbf{n})\nonumber\\
&+&g_{l}^{(N)}\left(k_{ln}^{(N)}r\right)\nabla
Y_{lm}(\mathbf{n})\Bigg]\label{eq:A_Mode_Again}\\
\nonumber\\
\mathbf{B}_{lnm}(\mathbf{x}) & = & -\left[l(l+1)\right]^{-1/2}\,
g_{l}^{(N)}\left(k_{ln}^{(N)}r\right)\,\mathbf{x}\nonumber\\
&\times&\nabla Y_{lm}(\mathbf{n})\label{eq:B_Mode_Again}\\
\nonumber\\
\mathbf{C}_{lnm,}(\mathbf{x}) & = &
k_{ln}^{(D)\,-1}\left[l(l+1)\right]^{-1/2}\nonumber\\
&&\Biggl\{\frac{l(l+1)}{r^{2}}\, Y_{lm}(\mathbf{n})\, g_{l}^{(D)}\left(k_{ln}^{(D)}r\right)\,\mathbf{x}\nonumber \\
&+&\frac{d}{dr}\left[rg_{l}^{(D)}\left(k_{ln}^{(D)}r\right)\right]\nabla Y_{lm}(\mathbf{n})\Biggr\},\label{eq:C_Mode_Again}
\end{eqnarray}
where $k_{ln}^{(D,N)}$ and $g_{l}^{(D,N)}\left(k_{ln}^{(D,N)}r\right)$ are the eigenvalues and eigenfunctions of the radially-separated Helmholz equation, Equation~(\ref{eq:Radial_Helmholtz}). 
The eigenfunctions  $g_{l}^{(D,N)}\left(k_{ln}^{(D,N)}r\right)$ are therefore the spherical Bessel functions.
Here, $D$ stands for the Dirichlet boundary
conditions and $N$ for the Neumann boundary conditions, $r\equiv |\mathbf{x}|$, $\mathbf{x}=r\mathbf{n}$, and the $Y_{lm}(\mathbf{n})$ are the usual spherical harmonic
functions. 

Note that since $\nabla Y_{lm}(\mathbf{n})$ and 
$\mathbf{x}\times\nabla Y_{lm}(\mathbf{n})$ are purely tangential 
vectors, the radial and tangential directions are explicitly 
separated in Equations~(\ref{eq:A_Mode_Again})--(\ref{eq:C_Mode_Again}). 
The set of 
modes, $\mathbf{A}_{nlm}$, is purely irrotational while the other 
two ($\mathbf{B}_{nlm}$ and $\mathbf{C}_{nlm}$) are purely 
solenoidal. 

A general momentum density vector field, $\mathbf{m}(\mathbf{x})$,
within a spherical shell ($R_{1} < r < R_{2}$)
can be uniquely decomposed into the VSH modes $\mathbf{A}_{nlm}$, $\mathbf{B}_{nlm}$ and $\mathbf{C}_{nlm}$ as follows:
\begin{eqnarray}
\mathbf{m}(\mathbf{x})&=&\sum_{lnm}\big[a_{lnm}\mathbf{A}_{lnm}(\mathbf{x})+b_{lnm}\mathbf{B}_{lnm}(\mathbf{x})\nonumber\\
&+& c_{lnm}\mathbf{C}_{lnm}(\mathbf{x})\big],\label{eq:Expansion}
\end{eqnarray}
where
\begin{eqnarray}
a_{lnm} & = & \int_{\Sigma}d^{3}\mathbf{x}\,\mathbf{m}(\mathbf{x})\cdot\mathbf{A}_{lnm}^{*}(\mathbf{x})\label{eq:A_Coefficients}\\
b_{lnm} & = & \int_{\Sigma}d^{3}\mathbf{x}\,\mathbf{m}(\mathbf{x})\cdot\mathbf{B}_{lnm}^{*}(\mathbf{x})\label{eq:B_Coefficients}\\
c_{lnm} & = & \int_{\Sigma}d^{3}\mathbf{x}\,\mathbf{m}(\mathbf{x})\cdot\mathbf{C}_{lnm}^{*}(\mathbf{x}).\label{eq:C_Coefficients}
\end{eqnarray}

While a field $\mathbf{m}(\mathbf{x})$ can be generally decomposed
in this way, not all the information thus generated is physically interesting.
The complex phase is not descriptive of scale structure.  Furthermore,
in the absence of a preferred direction, the values of components labeled by $m$
indices are a happenstance of choice of orientation -- they
shuffle among themselves while leaving $l$ and $n$ invariant under rotations. 
Therefore, in circumstances where we expect the power distribution among modes
to be spherically-symmetric, and where we are only concerned with scale structure, we may focus on describing
the scale structure of $\mathbf{m}(\mathbf{x})$ by computing the three rotationally-invariant spectra
\begin{eqnarray}
\alpha_{ln} & \equiv & \sum_{m}\left|a_{lnm}\right|^{2}\label{eq:alpha}\\
\beta_{ln} & \equiv & \sum_{m}\left|b_{lnm}\right|^{2}\label{eq:beta}\\
\gamma_{ln} & \equiv & \sum_{m}\left|c_{lnm}\right|^{2}.\label{eq:gamma}
\end{eqnarray}
If $\mathbf{m}(\mathbf{x})$ is from a simulation, these spectral
integrals may be obtained by straightforward quadrature summations
over the domain mesh. Evidently, in these spectral distributions,
the $l$ index is informative with respect to angular scales, whereas
the $n$ index is informative with respect to radial scales.

Since the $\mathbf{A}_{nlm}$ are purely irrotational, and therefore 
affect $\partial\rho/\partial t$, whereas the $\mathbf{B}_{nlm}$ and 
$\mathbf{C}_{nlm}$ modes are purely solenoidal, and have no effect 
on $\partial\rho/\partial t$, when analyzing existing momentum 
fields from nearly-stable stratified flows, one should expect to 
find that the magnitude of the $\alpha_{ln}$ spectrum is smaller 
than the magnitudes of the $\beta_{ln}$ and $\gamma_{ln}$ spectra.  We confirm this expectation
using simulations in the next section.

\section{NUMERICAL EVALUATION OF VSH}\label{sims}

 Here we discuss the numerical methods that we use
to calculate VSH power spectra from the output of multi-dimensional
simulations of convective velocity fields. Implementation of the numerics
can be run either as a stand-alone code, given momentum density data in spherical coordinates
for a convective shell exported by a simulation, or applied within 
a hydrodynamics code, like {\it FLASH}, at runtime. 

The first thing to consider prior to the calculation of VSH is determination of the
radii $R_1$ and $R_2$ that bound the spherical shell $\Sigma$, at each time
at which the decomposition is to be performed.
These radial boundaries must be set so that the boundary
conditions of Equation~(\ref{eq:MomBC}) are satisfied as well as possible. 
This can be formally done by plotting the spherical average of the radial component of the momentum density, $m_{r}(r)$,
in the region where convective instability due to shell
burning is active.  The boundaries can then be chosen to be the locations where $m_{r}$ is as
close to zero as possible. Note that in reality $m_{r}$ is not going to be exactly equal to zero because
the convective boundaries themselves may be dynamical \citep{2007ApJ...667..448M}.
This inspection will provide the first inputs needed for the calculation of VSH: the inner and the outer
convective shell boundaries $R_{1}$ and $R_{2}$. 

The next input needed is the shortest physical length $\lambda_{r}$ to be resolved by the VSH decomposition. 
Of course, $\lambda_{r}$ needs to be larger than the finest resolution length of the discrete simulation mesh
within the convective shell. By counting nodes of the mode functions, it is not difficult
to show that for a given $\lambda_{r}$ the
maximum number of $n$-modes for all three VSH sets is given by the integer part of:
\begin{equation}
n_{\mathrm{max}} = \frac{2(R_{2}-R_{1})}{\lambda_{r}},\label{eq:nmax}
\end{equation}
and the maximum number of $l$-modes by the integer part of:
\begin{equation}
l_{\mathrm{max}} = \frac{\pi(R_{1}+R_{2})}{2 \lambda_{r}}.\label{eq:lmax}
\end{equation}
In a specialized case, where $\Delta R/R$ is
extremely small (not the case here), one could consider a different
choice to conform to the extreme BC. 
The total number of modes is:
\begin{equation}
N_{\mathrm{total}} = (n_{\mathrm{max}}+1)(l_{\mathrm{max}}+1)^{N_{\mathrm{D}}-1},\label{eq:ntotal}
\end{equation}
where $N_{\mathrm{D}}$ the dimensionality of the simulation. The dependence on $N_{\mathrm{D}}$ reflects
the fact that in 2D with azimuthal symmetry, only the $m=0$ mode is non-zero,
whereas in 3D there are $2l+1$ modes indexed by $m$ for each $l$. 

Given these inputs all operations necessary to calculate the VSH power spectra can be performed,
as described in the previous section. These operations involve the numerical evaluation 
of 
\begin{itemize}
\item The eigenvalues $k_{ln}^{(D,N)}$ for each combination $l,n$ and each choice
of boundary condition $D,N$ (solutions of Equations~\ref{eq:Root_Eqn} and \ref{eq:NBC_Eqn2});
\item The value of the function $Y_{lm}(\mathbf{n})$, and its gradient,
for each combination $l,m$ in every mesh zone in the shell;
\item The values of the spherical Bessel functions of first
and second kind and their derivatives for each combination $l,n$ in every mesh zone in the shell. 
\end{itemize}
The three VSH sets of mode functions for all modes up to $l_{\mathrm{max}}$ and $n_{\mathrm{max}}$ are built up, and spectra
of the given momentum field are obtained by summation over the domain mesh.

We have incorporated the operations necessary for the calculation of VSH into {\it FLASH} and
have done performance tests yielding information on the computational expense of calculating VSH
power spectra. The typical performance for calculations run on the Argonne Leadership Computational
Facility (ALCF) {\it Mira} supercomputer (with 1.6~GHz cores) is 
$1.6535 \times 10^{-7}$~core-hours per mesh zone per VSH mode.  
For the VSH decomposition
of the 3D oxygen-burning shell simulation discussed in \S\ref{sims_3D}, the number of mesh zones in the shell is $N_{\mathrm{z,sh}} =$~73,701,376, and the total number of modes is $N_{\mathrm{total}}=$~108.  Using
512 nodes (with 16 cores per node), the wall-clock time of the computation was 9.64~min, for a total of 1316.1~core hours used.

\subsection{{\it Calculation of VSH power spectra.}}\label{2DSpectra}

\begin{figure}
\begin{center}
\includegraphics[angle=-90,width=3.25in,trim= 1.in 0.25in 0.5in 0.75in,clip]{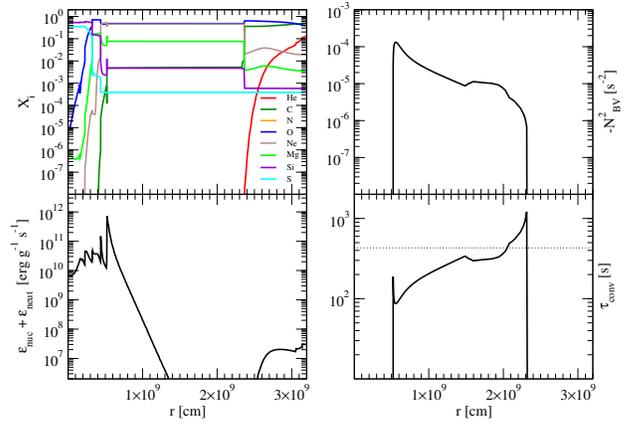}
\caption{Composition ({\it upper left panel}), total specific energy generation rate, 
$\epsilon_{nuc}+\epsilon_{neut}$ ({\it lower left panel}), convection coefficient ({\it upper right panel})
and convective turnover timescale ({\it lower right panel}) of the shell oxygen burning 15-$M_{\odot}$ model
at the time of mapping to {\it FLASH}. The dotted horizontal line indicates the time
$t =$~430~s when the simulation was stopped. This is the stage where
we perform the VSH decomposition.\label{Fig:2D_Model_Properties}}
\end{center}
\end{figure}

In this section we apply VSH decomposition to a 2D azimuthally-symmetric simulation of a convective oxygen-burning shell in an evolved
solar-metallicity 15-$M_{\odot}$ star. 
The 15-$M_{\odot}$ model was evolved from the Zero Age Main Sequence (ZAMS) up to the point where a predominantly Si/S core
is embedded in an O-burning shell, using the stellar evolution code {\it MESA} version 5596 
\citep{2011ApJS..192....3P, 2013ApJS..208....4P}. At this point the age of the star was 
$1.2 \times 10^{7}$~years, and 60\% of the oxygen fuel available for fusion was exhausted.
This phase is similar to that chosen by \citep{2007ApJ...667..448M}, who use a different stellar mass, however. 
{\it MESA} was run with the mass-loss prescriptions of \citet{1988A&AS...72..259D} and
\citet{2001A&A...369..574V}, the HELM EOS \citep{2000ApJS..126..501T}, 
the Schwarzschild criterion for convection and the approx19 nuclear reaction network \citep{1999ApJS..124..241T}. 
The final model has 1371 grid points, which yields good resolution of the convective flow in the O-burning shell. 
Figure \ref{Fig:2D_Model_Properties} shows the distribution of the specific nuclear energy generation rate, the convective overturn
time-scale, the square of the Brunt-Vaisala frequency, $N_{BV}^{2}$, and the composition of the final {\it MESA} model. For an estimate
of the convective overturn time-scale we take $\tau_{\rm conv} \simeq
1/\sqrt{N_{BV}^{2}}$. 

\begin{figure}
\begin{center}
\includegraphics[angle=0,width=3.55in,trim= 0in 1in 1.in 0.5in, clip]{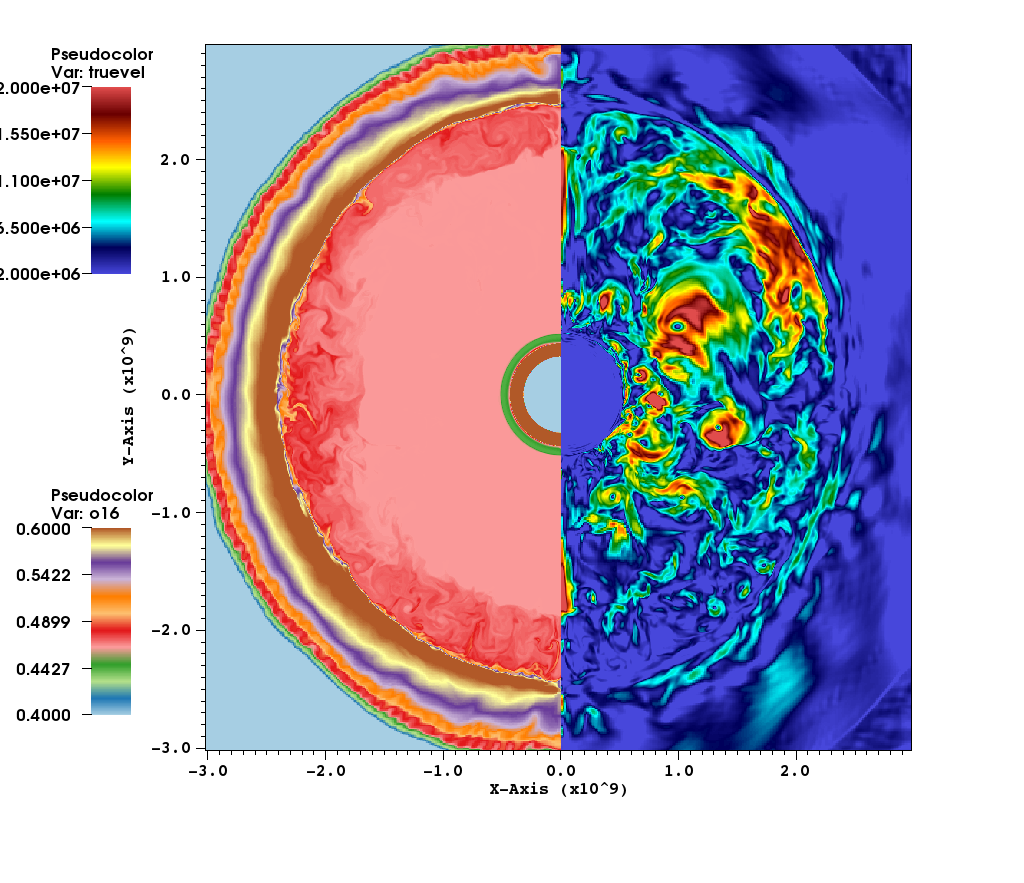}
\caption{$^{16}$O mass fraction ({\it left panel}) and velocity magnitude ({\it right panel}) of
the shell oxygen burning 15-$M_{\odot}$ model at $t =$~430~s after mapping to {\it FLASH}.\label{Fig:2D_Model_Viz}}
\end{center}
\end{figure}

\begin{figure}
\begin{center}
\includegraphics[angle=-90,width=3.25in,trim= 0in 0.3in 0.5in 0.5in, clip]{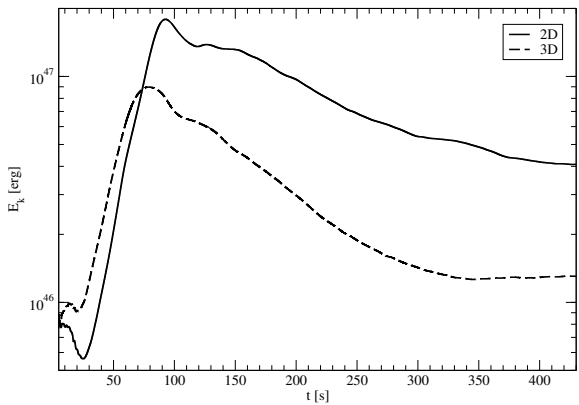}
\caption{Evolution of the total kinetic energy for the 2D (solid curve) and
the 3D (dashed curve) shell O-burning simulations.\label{Fig:TKE_evol}}
\end{center}
\end{figure}

This final model was then mapped to the 2D cylindrical Adaptive Mesh Refinement (AMR) grid of {\it FLASH}, version 4.0. 
The simulation was carried out in a cylindrical domain with maximum radial extent 
equal to $10^{10}$ cm and maximum/minimum vertical extents $\pm10^{10}$ cm, which easily contains the entire convective
O-burning shell. The total simulated time was $\sim$800~s
with 8 levels of refinement yielding a maximum resolution of $\sim$8~km. 
The levels of refinement were concentrically nested to yield an effective angular resolution of $\lesssim0.5^\circ$.
Within
the simulated timescale, convection over most of the oxygen-burning shell 
had undergone $\sim$~1-2 convective turnover timescales
(see lower right panel of Figure \ref{Fig:2D_Model_Properties}). 
The {\it FLASH} simulation was run
on the Texas Advanced Computing Center (TACC) {\it Stampede} supercomputer. 
Figure \ref{Fig:2D_Model_Viz} shows a snapshot of the $^{16}$O mass fraction and the
velocity magnitude in the oxygen shell after 430~s of simulation. 
By inspection of the net kinetic energy evolution of the
simulation, we find that the initial mild hydrodynamic transients caused
by mapping 1D {\it MESA} profiles to the 2D {\it FLASH} AMR grid
have ceased well before this time.
The initial transient manifests as a purely radial pulsation as
 the spherically-symmetric models settles onto the new grid.
To better illustrate these points, we show the evolution of the total kinetic
energy in Figure \ref{Fig:TKE_evol} for the 2D and the 3D shell O-burning simulation
we study in this work. The initial transient causes a peak in the kinetic energy just before 100~s in both 2D and 3D.  
By the end of the simulations, the radial pulses have left the domain and the kinetic energy is dominated by 
convective motions in the burning shells that have approximately saturated in strength.

\begin{figure}
\begin{center}
\includegraphics[angle=-90,width=3.25in, trim= 1.in .25in .5in 1in,clip]{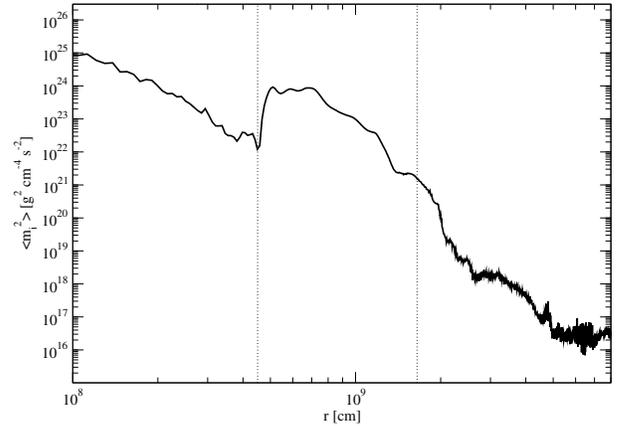}
\caption{Distribution of the radial component of $\langle m^{2} \rangle$ in radial bins for the 2D O-shell burning simulation at
430~s (black curve) and its realization (red curve). The vertical dotted lines
correspond the inner ($R_{1} = 0.45 \times 10^{9}$~cm) and the outer
($R_{2}= 1.65 \times 10^{9}$~cm)
radii chosen for the shell to be decomposed into VSH.\label{Fig:2D_Mr}}
\end{center}
\end{figure}

As a preliminary step to performing a VSH decomposition of the data, we determine the
radial boundaries of the 2D O-burning shell that best satisfy the boundary
conditions discussed in \S\ref{vsh}. To do so, we assign zones to radial bins, and
average the $m_{i}^{2}$
(where $i = r,\theta,\phi$) in each bin,  weighted by the volume of each zone:
\begin{equation}
\langle m_{i}^{2} \rangle_{k} = \frac{\sum_{q} m_{i}(k,q)^{2} dV(k,q)}{\sum_{q} dV(k,q)},
\end{equation}
where the index $k$ refers to a specific radial bin, and the index $q$ to
a zone within that bin.
Inspection of the binned momentum density data within the convective oxygen burning shell at $t =$~430~s 
leads us to choose shell boundaries at $R_{1} = 0.45 \times 10^{9}$~cm and $R_{2} = 1.65 \times 10^{9}$~cm (Figure \ref{Fig:2D_Mr}).
We then calculate the VSH modes using $\lambda_{r} = 1 \times 10^{8}$~cm (1/12 of the shell width). Using
Equations (\ref{eq:nmax})-(\ref{eq:ntotal}), this corresponds to $l_{\mathrm{max}}=$~32, $n_{\mathrm{max}}=$~24 and $N_{\mathrm{total}} =$~825.

\begin{figure}
\begin{center}
\includegraphics[angle=-90,width=3.25in, trim= 1.25in .5in .75in 1.25in,clip]{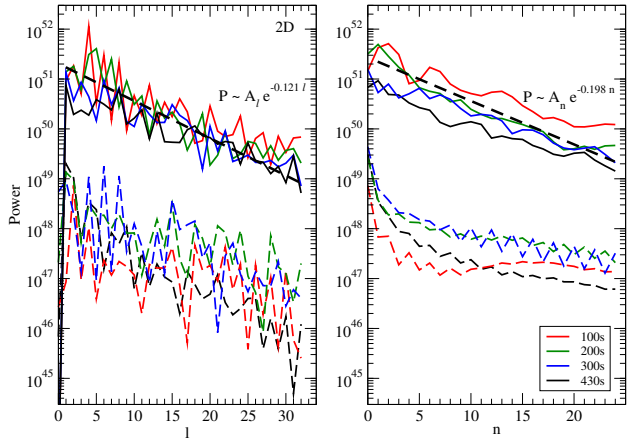}
\caption{Evolution of the reduced power spectra of the 2D shell O-burning simulation 
for $l$ (left panel) and $n$ (right panel). Dashed curves are for the $\alpha$ (irrotational) and solid curves
for the $\gamma$ (solenoidal) spectra. The legend indicates the time
(in seconds) for each spectrum. The orange curves show a  time-averaged
spectrum over the phases considered. Both the l- and n-spectra have similar slope in
all epochs respectively, that can be well represented by exponential laws.\label{Fig:2D_Spectra}}
\end{center}
\end{figure}

\begin{figure}
\begin{center}
\includegraphics[angle=-90,width=3.25in, trim= 1.25in .4in .75in 1.25in,clip]{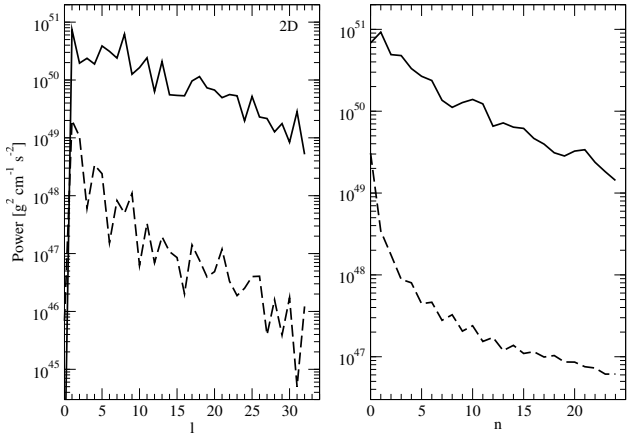}
\caption{The final reduced power spectrum of the 2D shell O-burning simulation at 430~s
for $l$ (left panel) and $n$ (right panel). Dashed curves are for the $\alpha$ (irrotational) and solid curves
for the $\gamma$ (solenoidal) spectra. \label{Fig:2D_430sSpec}}
\end{center}
\end{figure}

Figure \ref{Fig:2D_Spectra} shows the temporal evolution of the reduced power spectra in $l$ (left panel)
and $n$ (right panel). The reduced power spectra are essentially
summations of the full spectra over either $l$ or $n$: 
$\alpha_{l}^\prime\equiv\sum_{n} \alpha_{nl}$, $\alpha_{n}^{\prime\prime}\equiv\sum_{l} \alpha_{nl}$, and similarly for the solenoidal modes.
Due to the lack of $m \neq$~0 modes and the
azimuthal symmetry of the 2D data all the $\mathbf{B}_{lnm}$ modes have zero coefficients,
that is $\beta_{nl}=0$. 
The total power (summed over all $l$ and $n$) of the irrotational mode is $4.128 \times 10^{49}$~g$^{2}$~cm$^{-1}$~s$^{-2}$
and that of the solenoidal mode $4.594 \times 10^{51}$~g$^{2}$~cm$^{-1}$~s$^{-2}$.
We therefore have $\alpha_{ln} < \gamma_{ln}$ as expected from the discussion in Section \ref{vsh}. 
A time-averaged spectrum over the four phases considered (100-430~s) is also shown in Figure \ref{Fig:2D_Spectra}
to illustrate variance of power over time in each wavenumber. While there is considerable variance about the mean 
value for each individual wavenumber, illustrating the chaotic nature of convection, 
the global characteristics of the spectra (e.g., the slopes) remain consistent.
The reduced VSH power spectrum of the convection simulation at $t=430$~s, shown in Figure
\ref{Fig:2D_430sSpec}, yields scale information
about flows in the convective oxygen shell. The $\gamma_{ln}$-spectrum begins its exponential decline at $l =$~8 and at $n =$~1. This means that the bulk of the momentum density power is in  characteristic angular scales of 
$\sigma_{\theta} = 7.3 \times 10^{8}$~cm and radial scales of $\sigma_{r} = 6.0 \times 10^{8}$~cm,
indicating that the size of the convective elements is roughly similar in the radial and angular directions. 
We also note that the power of the solenoidal modes in $l$ maintains
a constant slope in time that can be approximated by a declining exponential
of the form $P \propto \exp(-0.121 l)$. Same is the case for the power
of the solenidal modes in $n$ with corresponding declining exponential
law $P \propto \exp(-0.198 n)$. We emphasize that these power-law fits are only indicative
and we only use them to illustrate the general characteristics of the power spectra and not for realizations
of velocity fields. For this purpose we recover the phase information by random drawings as
discussed in Section~\ref{Realization}.

\subsubsection{{\it Orthonormality of modes.}}

\begin{figure}
\begin{center}
\includegraphics[angle=-90,width=3.25in, trim= 1.in .4in .75in 1.25in,clip]{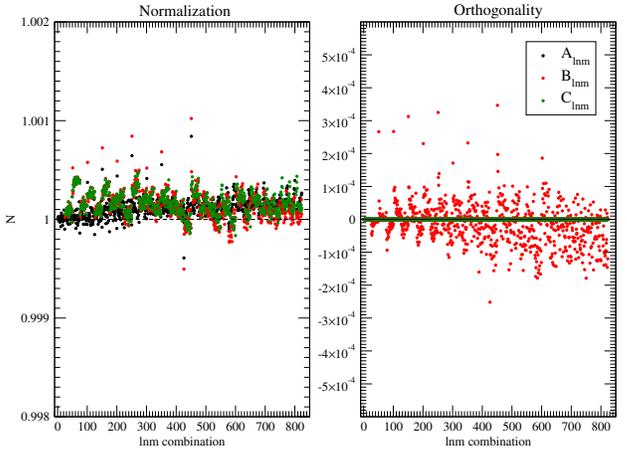}
\caption{Normalization (left panel) and orthogonality parameters (right panel) tests for the VSH modes using the data
of the 2D convective O shell simulation. This test illustrates that
the VSH modes we derive in the Appendix and use in this work are
properly orthonormal. \label{Fig:OrthoNormal}}
\end{center}
\end{figure}

For the VSH calculation presented in \S\ref{2DSpectra} we test the orthonormality of the VSH modes in order
to assess the consistency of our method, and to verify the correctness of the VSH code. For this we evaluate the output of 
Equations~(\ref{eq:A_Mode_Orthogonality}), (\ref{eq:B_Mode_Orthogonality}), (\ref{eq:C_Mode_Orthogonality}) and (\ref{eq:Mutual_Orthogonality}) of Appendix \ref{sec:appendix}
and verify that the results are either 0 (orthogonal)
or 1 (normalized). Our results are illustrated in Figure \ref{Fig:OrthoNormal}.
It can be seen that all three sets of modes have normalization factors very close to unity.
Also, the three sets of VSH modes are orthogonal to each other. 

\begin{figure}
\begin{center}
\includegraphics[angle=-90,width=3.25in, trim= 1.in .25in .5in 1in,clip]{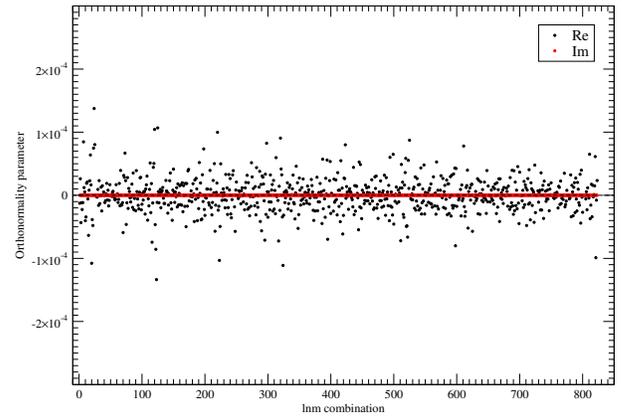}
\caption{Mutual orthonormality parameter for the $A_{010}$ mode against all $A_{lnm}$ modes. Filled black circles
are for the real part and filled red circles for the imaginary part of the parameter. For clarity we do not expand
the y-axis up to unity but we wish to 
note that the parameter is very close to zero in all cases except for the case $l =$~0, $n =$~1, $m =$~0 for which
the real part is 1 indicating that the mode is properly normalized.\label{Fig:Mutual_Ortho}}
\end{center}
\end{figure}

To verify that different modes of the
same set are mutually orthogonal we use the $n =$~1, $l =$~0 mode and test its orthogonality
against all other modes. As can be seen in Figure \ref{Fig:Mutual_Ortho}, orthogonality is recovered in this case
as well.

\subsubsection{{\it Recovery of the original data.}}

\begin{figure*}
\begin{center}
\includegraphics[angle=0,width=7.4in,trim= 1in 2in 1in 1in, clip]{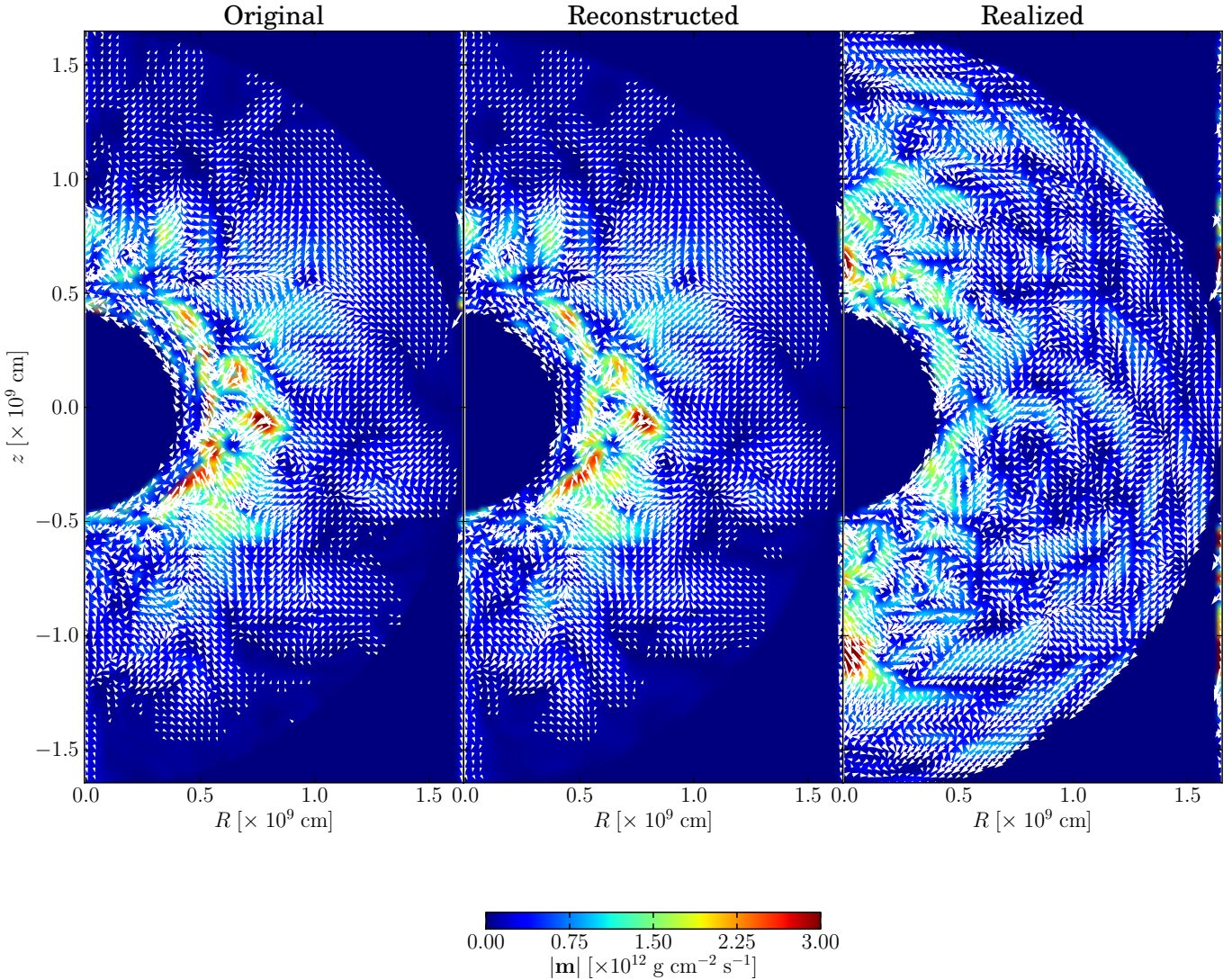}
\caption{{\it Left Panel}:  
Original momentum density field from the convective O-shell 2D simulation.
{\it Middle Panel}: Reconstructed field (Section \ref{Realization}).
{\it Right Panel}:  Realized momentum density field in the convective O-shell using the VSH power spectrum
obtained in Section \ref{2DSpectra}.\label{Fig:Realization}}
\end{center}
\end{figure*}

Another test of the VSH method is whether the original momentum density data can be recovered
accurately by using the decomposition of Equations (\ref{eq:Expansion})-(\ref{eq:C_Coefficients}). 
Using the  VSH decomposition of the convective oxygen-burning shell that we presented
in Section \ref{2DSpectra}, we reconstruct the momentum density data using Equation~(\ref{eq:Expansion}). 
Figure \ref{Fig:Realization} shows a comparison of the recovered (middle panel) versus the original (left panel)
momentum density field. 
For the two datasets
we calculate the L2-norm to assess the error of the reconstructed field using the
formula:
\begin{equation}
f = \sqrt{\frac{\int_{\Sigma} d^{3}\mathbf{x}\left[\mathbf{m}(\mathbf{x})-\mathbf{m^\prime}(\mathbf{x})\right]^{2}}
{\int_{\Sigma} d^{3}\mathbf{x}\left|\mathbf{m}(\mathbf{x})\right|^{2}}}.
\end{equation}
We find $f =$~0.0029 indicating a successful reconstruction of the data within errors (since $f <<$~1). 
Part of the residual deviation of the reconstructed momentum density
field from the original one is due to the truncation of the sum over modes at $l=l_{\mathrm{max}}$, $n=n_{\mathrm{max}}$.
Some of the residual is also attributable to the fact that at $R_{1}$ and $R_{2}$ the boundary conditions
are not perfectly satisfied (see Section \ref{sims}).

\subsection{{\it Realization of simulation data from VSH power spectra.}}\label{Realization}

An important application for the VSH framework presented here is the 
generation of non-spherically-symmetric initial conditions for CCSN simulations that include physically-motivated non-radial velocity fields.
Inclusion of non-radial velocity perturbations to otherwise
spherically-symmetric 
initial conditions has been shown to have an important qualitative
impact on CCSN 
simulations \citep{2013ApJ...778L...7C}.
This work, however, used very simplistic convolutions of sinusoids to perturb only the $\theta$-direction velocity of the initial model.
The scale and amplitude of these perturbations were chosen to resemble
realistic multi-dimensional simulations of convective 
burning in CCSN progenitors near collapse, but this approach is
ultimately insufficient to quantitatively capture the structure of 
convective velocity fields.
Our method of VSH decomposition provides a far superior means of
including non-radial velocity fields in the initial conditions 
for CCSN simulations.
As discussed above, the VSH framework is purely kinematic in the sense
that it includes no 
information about the physics that is producing the velocity field
being considered, 
but it does respect appropriate boundary conditions for convective shells in stars.
In this sections, we describe the process of realizing a pseudo-random velocity field from VSH spectra.

Under the assumption that the spectral distributions $\alpha_{ln}$, $\beta_{ln}$,
$\gamma_{ln}$ are known we can create corresponding realizations
of $\mathbf{m}(\mathbf{x})$. For this, we need to restore the
phase and the $m$-dependence by random number generation so as 
to turn the power spectra back into full VSH transforms.

In order to do this, we observe that since the data is real, the expansion 
coefficients in Equations~(\ref{eq:A_Coefficients})--(\ref{eq:C_Coefficients}) 
have the property $a_{lnm}^*=a_{ln\,-m}$, $b_{lnm}^*=b_{ln\,-m}$, 
$c_{lnm}^*=c_{ln\,-m}$ -- this is traceable to the relation 
$Y_{lm}(\mathbf{n})^*=Y_{l\,-m}(\mathbf{n})$ which is inherited by the mode 
functions $\mathbf{A}_{lnm}$, $\mathbf{B}_{lnm}$, $\mathbf{C}_{lnm}$. It follows
that for each mode, the process of extracting the spectral coefficients $\alpha_{ln}$,
$\beta_{ln}$, $\gamma_{ln}$ according to Equations~(\ref{eq:alpha})--(\ref{eq:gamma})
has elided the sign of the $m=0$ term (which is purely real) and the phases of
the $m>0$ terms.  The remaining $m<0$ terms are related to the $m>0$ terms by
complex conjugation.  We may therefore focus on realizing the truly independent $(2l+1)$ real terms
$m\ge 0$ for each mode, and obtain the $m<0$ terms by complex conjugation.

We may accordingly rewrite Equations~(\ref{eq:alpha})--(\ref{eq:gamma}) as follows:
\begin{eqnarray}
\alpha_{ln}&=&a_{ln0}^2+
2\sum_{m=1}^l\left[\textit{Re}\left\{a_{lnm}\right\}^2
+\textit{Im}\left\{a_{lnm}\right\}^2
\right]\label{eq:alpha_real}\\
\beta_{ln}&=&b_{ln0}^2+
2\sum_{m=1}^l\left[\textit{Re}\left\{b_{lnm}\right\}^2
+\textit{Im}\left\{b_{lnm}\right\}^2
\right]\label{eq:beta_real}\\
\gamma_{ln}&=&c_{ln0}^2+
2\sum_{m=1}^l\left[\textit{Re}\left\{c_{lnm}\right\}^2
+\textit{Im}\left\{c_{lnm}\right\}^2
\right]\label{eq:gamma_real}
\end{eqnarray}

Setting the components of a $(2l+1)$-dimensional vector $\mathbf{y}^{(\alpha,l,n)}$
by
\begin{equation}
[\mathbf{y}^{(\alpha,l,n)}]_p=\left\{
\begin{array}{c@{\quad:\quad}l}
\sqrt{\frac{1}{\alpha_{ln}}}a_{ln0}&p=0\\
\sqrt{\frac{2}{\alpha_{ln}}}\textit{Re}\left\{a_{lnp}\right\}&p=1,\ldots,l\\
\sqrt{\frac{2}{\alpha_{ln}}}\textit{Im}\left\{a_{ln\,p-l}\right\}&p=l+1,\ldots,2l,
\end{array}
\right.
\label{eq:x_vector}
\end{equation}
it is clear that Equation~(\ref{eq:alpha_real}) is satisfied when 
$|\mathbf{y}^{(\alpha,l,n)}|^2=1$, that is, when
$\mathbf{y}^{(\alpha,l,n)}$ is restricted to a unit hypersphere in $(2l+1)$
dimensions.  We may therefore realize 
a velocity distribution that respects the spectral distributions 
$\alpha_{ln}$ by constructing, for each $l,n$, a random vector $\mathbf{y}^{(\alpha,l,n)}$
on this sphere. 

Sampling such a random vector is a straightforward matter.  We sample 
$(2l+1)$ standard normal variables $\hat{y}_p$, $p=0,\ldots,2l$. We then 
compute the normalization $N^{2}\equiv \sum_{p=0}^{2l}\hat{y}_p^{2}$. Since 
the density function of the $(2l+1)$-dimensional multivariate normal 
distribution depends only on $N$, the normalized random variables 
$[\mathbf{y}^{(\alpha,l,n)}]_p=\hat{y}_p/N$ are distributed uniformly on the 
unit hypersphere in $(2l+1)$-dimensions, as required.  The realized values 
of $a_{lnm}$ may be easily recovered from Equation~(\ref{eq:x_vector}).  This 
procedure may obviously be repeated to obtain realizations of the spectra 
$\beta_{ln}$ and $\gamma_{ln}$. Of course, statistically-independent random 
number drawings should be performed for each of the three sets of modes.
Once all ($a_{lnm}$, $b_{lnm}$, $c_{lnm}$) full power spectrum coeffiecents
have been recovered, Equation~\ref{eq:Expansion} can be used to obtain the final realized momentum
density field.

We note that for 2D momentum density data the realization process is 
different since only the $m=0$ mode is present, which is only uncertain up 
to a sign. In this case the random variables are drawn from a uniform 
distribution instead of a normal one, and are assigned a positive or a 
negative unity value according to whether they fall in the (0-0.5] or the 
(0.5-1] range respectively.

We use the VSH power spectra calculated in Section \ref{2DSpectra} to produce a momentum density field
realization by the process described above. The realized momentum density field for the O shell can be
seen in Figure \ref{Fig:Realization} (right panel) where it is compared with the original momentum density
field (left panel). We observe that qualitatively, scales of macroscopic structures occurring in the original
simulation also recur in the realized field. 

It should be noted that the spectra
only bear scale information, but do not carry any location information.
This means that spatial structure such as stratification can be lost
when $\mathbf{m}(\mathbf{x})$ is analyzed into modes and re-realized from the
resulting spectrum. A stratified distribution
$\mathbf{m}(\mathbf{x})$ with a quiescent core and a convective outer
shell, spectrally-analyzed using whole-star modes and re-realized
as above, will in general produce new momentum distributions with the same mix
of scales as $\mathbf{m}(\mathbf{x})$, but with no core-shell structure.
If that structure must be preserved, it is necessary to divide the
star into as many concentric layers as required, and analyze each
layer separately.
Nevertheless, the initial VSH decomposition requirements on mass, momentum conservation and
boundary conditions (Equations~\ref{eq:MassCons}-\ref{eq:MomBC})
are already built in the method and the realization of phase
information does not violate them. 
The conservation of mass flux, as we stated it in \S~\ref{vsh}, is clearly visible in Figure~\ref{Fig:2D_Spectra}, 
where the divergence-bearing mode ($\alpha$) is clearly much smaller than the solenoidal modes.

More generally, realization by ``restoring the phases'' is a seemingly 
simple operation that is, however, fraught with physical significance.  The 
VSH decomposition itself is, of course, purely a kinematic device, and 
nothing about the spatial coherence scales of the flow, such as might be 
determined by convective turbulence structures associated with time-domain 
intermittency, is included into the method.  Depending on the distribution 
chosen to create randomly sampled phases, such flow structures can be 
realized or wiped out. Here we discuss a realization strategy in which the 
phases are all independent, identically distributed (IID) on the unit 
circle.  This strategy is fully incoherent, and definitely wipes out any 
coherent spatial structure. This issue is clearly illustrated by
considering the distribution of $\langle m^{2} \rangle$ for the realized
simulation as compared to the original data (red curve in Figure \ref{Fig:2D_Mr}).
Although the original and the realized radial components of the momentum density fields
have a different spatial distribution, the average integrated value over the convective shell
is very similar in the two.
We return to a discussion of this issue in \S\ref{disc_2}.

There is, as yet, no multi-dimensional simulation of the late stages of massive stellar evolution that ends at the point of core collapse in the literature.
There are simulations, however, of convective Si shell burning in massive stars with iron cores \citep[e.g.,][Couch et al., in prep.]{2011ApJ...733...78A}.
As compared with detailed 1D CCSN progenitor models \citep[e.g.,][]{2002RvMP...74.1015W}, these simulations typically make approximations such 
as reduced nuclear burning networks and replacement of the inner iron core by a boundary condition which make attaining core collapse difficult.
Furthermore, these multi-dimensional simulations should not be regarded as fully deterministic as two similar simulations in identical progenitors
 that were subjected to different small scale perturbations in order to seed convection may result in well-developed convective structures 
that are not identical in phase.
Until multi-dimensional CCSN progenitor models at the point of collapse are available, a step toward increased realism in 
CCSN simulations may be achieved be leveraging the VSH method we present here in conjunction with state-of-the-art 1D models.

The general procedure for constructing physically-motivated aspherical initial conditions for use in CCSN simulations would be as follows.
First, a suitable simulation of convective burning prior to core collapse in the progenitor of interest is obtained.
The structure of the velocity field in, e.g., the Si-burning shell surrounding the iron core is analyzed with the VSH 
framework and the spectra, Equations (\ref{eq:alpha}-\ref{eq:gamma}), are constructed.
From the VSH spectra, we generate realizations of the velocity field using the procedure described above in this section 
which are applied to the Si shell of the 1D collapse model.
This yields a vector velocity field with scale and structure statistically equivalent to the original convective burning simulation, in terms of the vector spherical harmonic information, but with a randomized phase.
A goal of future work is to study whether or not the results of CCSN simulations are sensitive to the phase information 
of the convective velocity field in the Si shell at the time of collapse by conducting multi-dimensional CCSN mechanism 
simulations with several VSH-realized initial velocity fields that differ only in the random seed number of the phasing. 
A short-coming of this approach is the possibility for a small divergence of the evolution between the multi-dimensional 
convective burning simulation and that of the detailed 1D model.
For our purposes, however, what matters most is accurately capturing the strength and scale of 
the convection in the moments prior to core collapse.
We contend that the method sketched here should do just that and will result in initial conditions for 
CCSN that are more physically representative of real stars than anything that has been done before.

\section{APPLICATION TO THREE-DIMENSIONAL SIMULATION DATA}\label{sims_3D}

\begin{figure}
\begin{center}
\includegraphics[angle=0,width=3.25in,trim= 0in 1in 3in 6in,clip]{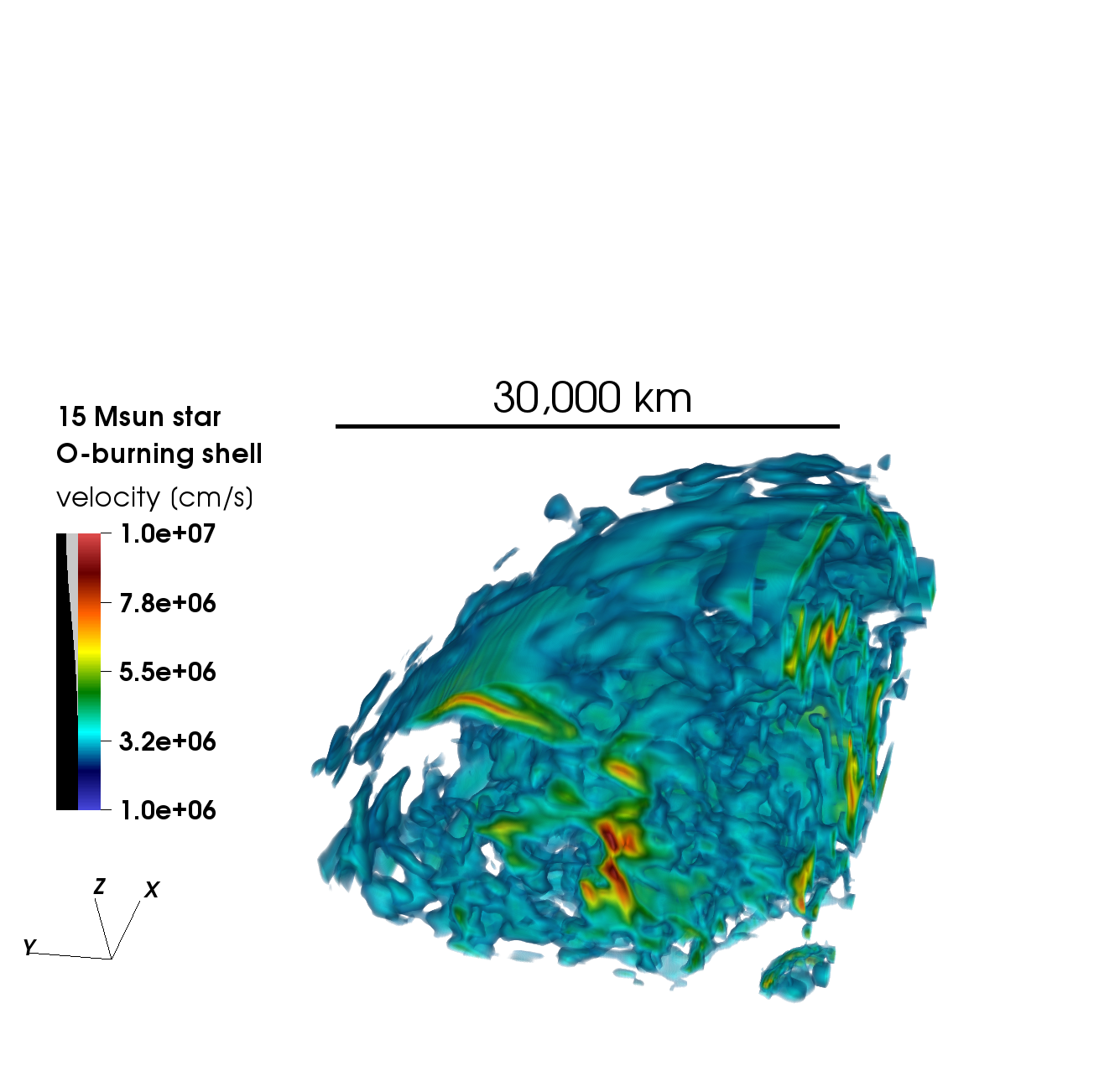}
\caption{Volume rendering of the velocity magnitude of the 15-$M_{\odot}$ convective
oxygen shell burning 3D simulation at 430~s.  The coordinate triad is shown in the bottom left.  
The relatively quiescent Si/S core is in the bottom right of the rendering. \label{3D_Model_Viz}}
\end{center}
\end{figure}

 To apply the VSH decomposition method on 3D data we ran a 3D FLASH
 simulation featuring a 15-$M_{\sun}$ star O-burning shell
on the Argonne Leadership Computing Facility (ALCF) {\it Intrepid}
supercomputer using the same progenitor as for the 2D study. 
 We used 8$^{3}$-zone blocks in a $10^{10}$~cm octant simulation box with reflective boundaries at the coordinate planes.
The simulation used Cartesian coordinates and Adaptive Mesh Refinement (AMR), with a maximum resolution of $\sim$~16~km.
It was evolved for 430 s. Figure \ref{3D_Model_Viz}
shows a snapshot of the velocity magnitude at this latest simulated
phase. Characteristic convective velocities of a few
$10^{6}$-$10^{7}$~cm~s$^{-1}$ are
obtained, which are similar to the ones seen in the 2D simulation.

In order to obtain the correct VSH power spectra we constructed the
full 4$\pi$ steradian momentum density data using reflection symmetry
about the three planes x-y, x-z and y-z. The radial coordinate $r$ and
$m_{r}$ remain unaltered in all eight octants while the angular
components of $\mathbf{m}$ had either the same or an opposite sign
depending on their specific octant. Of course, to recover the $\phi$
coordinates in the other octants we added multiples of $\pi/2$ while
$\theta$ was the same for all octants in the northern hemisphere and
$\theta + \pi/2$ for the southern hemisphere. Therefore the total
number of zones in the O-burning shell for which we applied the VSH
decomposition is eight times the number in the original octant
simulation.

To compute power spectra with the large number of zones and modes in the 3D case, we implemented
a parallelized computation of VSH in
{\it FLASH}. Prior to doing so, we 
re-weighted the AMR Morton space-filling curve \citep{Warren1993} to assign more
weight to blocks residing in the O-burning shell,
so that most processing power is focused on parts of the domain where it is
needed. To reduce computational cost we also de-refined the AMR
data outside the O-burning shell prior to
applying the VSH decomposition. These three processes of
parallelization, de-refinement of unused data and re-weighting of the
Morton curve allowed us to efficiently compute 3D VSH spectra from within FLASH
using the ALCF {\it Mira} supercomputer.

\begin{figure}
\begin{center}
\includegraphics[angle=-90,width=3.25in, trim= 1.in .3in .5in 1.25in,clip]{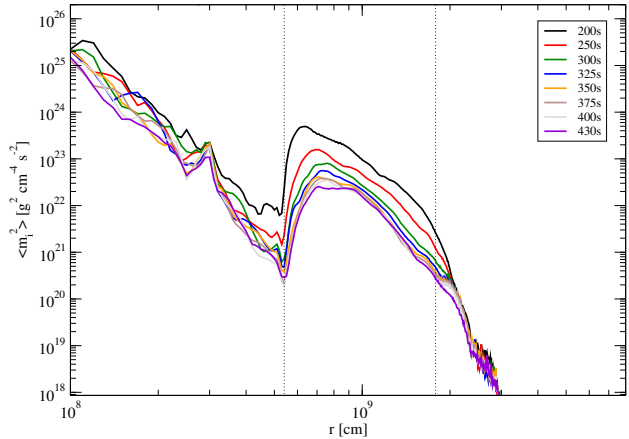}
\caption{Distribution of the radial component of $\langle m^{2} \rangle$ in radial bins for the 3D O-shell burning simulation at
different times indicated in the inset. The vertical dotted lines
correspond the inner ($R_{1} = 0.5405 \times 10^{9}$~cm) and the outer
($R_{2}= 1.7841 \times 10^{9}$~cm)
radii chosen for the shell to be decomposed into VSH.\label{Fig:3D_Mr}}
\end{center}
\end{figure}

First, we determine the radial limits of the O-burning shell. The binned
profiles of $m_{i}^{2}$ for several epochs are shown in Figure
\ref{Fig:3D_Mr}. The data shows that the radial component $\langle m_r^2 \rangle$
has a minimum at $R_{1} = 0.5405 \times 10^{9}$~cm. We choose this value
to be the inner boundary of the convective shell to be decomposed into
VSH. For the outer boundary we investigate two values: a value that
corresponds to the same $\langle m_{r}^{2} \rangle$ as the inner boundary ($R_{2} =
1.7841 \times 10^{9}$~cm) and a larger value for which $\langle m_{r}^{2} \rangle$ has 
even more significantly declined ($R_{2} = 3 \times 10^{9}$~cm) in
order to verify the robustness of the spectra under different choices of outer boundary. 

\begin{figure}
\begin{center}
\includegraphics[angle=-90,width=3.25in, trim= 1.25in .4in .5in 1.25in,clip]{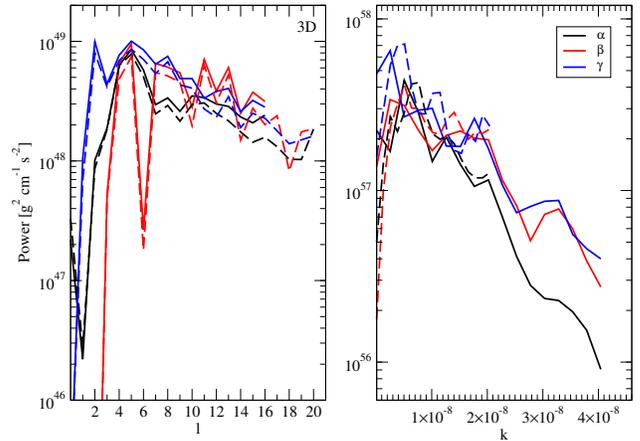}
\caption{Reduced VSH power spectra in $l$ (left panel) and $k$ (right
  panel) for the 3D O-shell burning simulation at 430~s. The solid
  curves are for the original choice of $R_{2} = 1.7841 \times
  10^{9}$~cm while the dashed curves for a longer outer radius ($R_{2}
= 3 \times 10^{9}$~cm). Black curves correspond to the irrotational ($\alpha$) mode, 
red curves to the first solenoidal ($\beta$) mode and blue curves to the second 
solenoidal ($\gamma$) mode.\label{Fig:3D_Spectra_AltR2}}
\end{center}
\end{figure}

\begin{figure}
\begin{center}
\includegraphics[angle=-90,width=3.25in,trim= 1.25in .4in .75in 1.25in,clip]{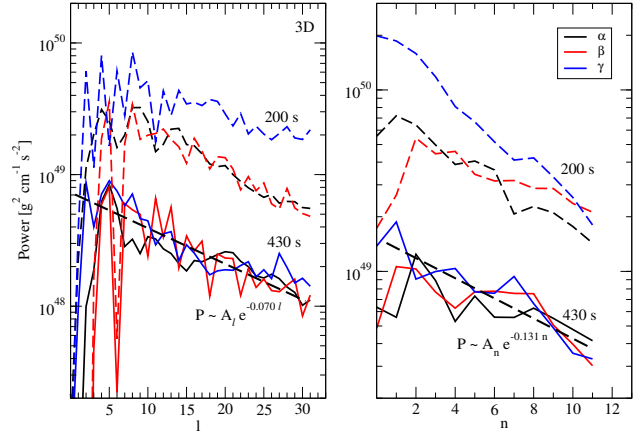}
\caption{Reduced VSH power spectra in $l$ (left panel) and $n$ (right
  panel) for the 3D O-shell burning simulation at 200~s (dashed
  curves) and 430~s (solid curves). The meaning of the different colors
is the same as in Fig~\ref{Fig:3D_Spectra_AltR2}. The thick lines show the time-averaged
spectra over the two phases considered.\label{Fig:3D_Spectra_Evolution}}
\end{center}
\end{figure}

We then run {\it FLASH} with the VSH implementation discussed above on
512 nodes (8192 cores) on {\it Mira} for 11.2 hours for a total of
91750 core hours per run. For these calculations we used $l_{\mathrm{max}} =$~20 and
$n_{\mathrm{max}} =$~16 for a total of $N_{\mathrm{total}} =$~7497 VSH modes calculated.
Figure \ref{Fig:3D_Spectra_AltR2} shows the resulting reduced VSH power spectra in the case
of the smaller (solid curves) and the larger (dashed curves) $R_{2}$
at 430~s. To properly compare the reduced VSH power spectrum
in $n$ given the different shell sizes we express it in terms
of the wavenumber $k$ in the right panel of Figure \ref{Fig:3D_Spectra_AltR2} using
$k = n \pi/(R_{2}-R_{1})$ and $dP/dk = [(R_{2}-R_{1})/\pi]  dP/dn$. We
find that choosing a larger outer radius does not significantly change the
resulting VSH power spectra. We also show the evolution of the power spectra
between 200~s and 430~s in Figure \ref{Fig:3D_Spectra_Evolution} as well as a time-averaged
spectrum over these two phases (thick curves). While the slopes of the spectra remain consistent
over time, considerable variance is observed for individual wavenumbers in accord with the 
2D results.
We see that the solenoidal $\mathbf{C}$ modes
dominate over the irrotational modes while the solenoidal $\mathbf{B}$ modes
are comparable to the irrotational modes. It is therefore still the case
that more power goes to the solenoidal modes, but the solenoidal/irrotational power ratio
is smaller than was the case in the 2D simulations.
More specifically at 430~s in the 3D case, the total power of the two solenoidal modes together
is $1.874 \times 10^{50}$~g$^{2}$~cm$^{-1}$~s$^{-2}$ while that of the irrotational
mode $7.773 \times 10^{49}$~g$^{2}$~cm$^{-1}$~s$^{-2}$. We speculate that the
increased importance of the irrotational flow with respect to the solenoidal
flow is at least in part due to the physically unrealistic octant boundary conditions,
which create radial-flow artifacts at the boundary planes.
The lower resolution of the 3D simulation (16~km versus 8~km for the 2D simulation) may also
contribute to the difference in power ratios.

Since the choice of a larger $R_{2}$ does not alter the VSH power
spectra significantly we adopt the smaller value $R_{2} = 1.7841
\times 10^{9}$~cm to calculate the final 3D convective O-burning shell
VSH power spectra at two epochs: 200~s and 430~s for
$l_{\mathrm{max}} =$~15 and $n_{\mathrm{max}} =$~11. The result is shown in Figure~\ref{Fig:3D_Spectra_AltR2}. 
We find that the slope of the solenoidal power spectrum in $l$
does not change significantly over this 230~s period in agreement with 2D results. 
However, the 430~s spectrum in this case peaks at $l=$~5 for both
solenoidal modes and at $n=$~2. These peaks corresponds to
characteristic angular scales $\sigma_{\theta,\phi} = 1.22 \times
10^{9}$~cm and radial scales $\sigma_{r} =4.15 \times 10^{8}$~cm
indicating that the convective elements tend to be oblately elongated in
the tangential directions. Finally, these
findings suggest that in 3D, convection in the O-burning shell moves more
power to smaller radial scales but larger angular scales than in 2D.
We also evaluated the exponential law slopes for the 3D power spectra at $t =$~430~s
and recovered $P \propto \exp(-0.070 l)$ and $P \propto \exp(-0.131 n)$ accordingly
for the $l$- and $n$-reduced spectra, indicating that the slope in $l$ is flatter
than in the 2D case while the slopes in $n$ are consistent.

\section{SUMMARY AND CONCLUSIONS}\label{disc_2}

In this paper we present the method of Vector Spherical Harmonics (VSH)
decomposition to characterize convective stellar velocity fields by
calculating power spectra that yield information about the
scales where convection has the most power, the distribution of convective
power over those scales, and the shape of typical
convective elements. We discussed our numerical implementation of the method
and used it to calculate power spectra given a momentum density
field from a simulation and the inner and outer radii of the
convective region that we wish to decompose into VSH. We also
discussed the techniques used to implement this method for use in large scale
3D simulations  that can be used to analyze
3D {\it FLASH} AMR hydrodynamics simulations
of stellar convection using large parallel computing resources, and can also be
ported to other hydrodynamics codes. 

We highlight the importance of properly characterizing
multi-dimensional convection in pre-SN stars prior to initiating
CCSN simulations. As previous studies have
shown \citep{2007ApJ...667..448M,2011ApJ...733...78A,2011ApJ...741...33A} the
one-dimensional mixing-length theory for convection has difficulty in
capturing the true nature of stellar convection in
convective thin shell burning such as is expected in massive evolved
stars. Vigorous convection in the oxygen shell during the core Si/S
burning phase and in the silicon shell in the hours prior to iron
core-collapse may significantly alter the structure of the progenitor
star (Couch et al. 2014, in preparation). It can
also affect the core-collapse dynamics and susceptibility to a
successful explosion in 3D \citep{2013ApJ...778L...7C}. 
In addition, the
large scale plumes that develop in realistic multi-dimensional convection in
oxygen-burning shells can produce gravity waves
that can lead to episodic mass-loss prior to the explosion,
which affects the initial circumstellar environment within which the
SN takes place \citep{2012MNRAS.423L..92Q, 2014ApJ...780...96S}.

These results highlight that the core-collapse SN problem
is an initial value problem, and that realistic 3D convective stellar
velocity fields must be imposed on the progenitor models as initial conditions
to hydrodynamic simulations \citep{2013ApJ...778L...7C}. We can use the VSH method
to produce physically-motivated realizations of stellar velocity fields in the
convective shells of SN progenitors using the derived VSH power
spectra from multi-dimensional simulations, yielding more realistic, non-spherical initial conditions.

We have calculated the time evolution of reduced power spectra
in both the angular ($l$-modes) and the radial ($n$-modes) scales for
2D and 3D convective oxygen shell burning of a 15-$M_{\odot}$ star.
We have also used the 2D analysis and results to verify and test the
predictions and accuracy of our method and to produce a realized
2D velocity field for the convective oxygen shell that bears the same
scale information as the original field.

In connection with realization, however it is important to emphasize that 
the realization strategy that we implemented in \S\ref{Realization} -- with IID 
uniform phases -- is in a certain sense naive, at least for some purposes.  
The point is that the VSH decomposition is a purely kinematic device, which 
does not incorporate any dynamical information a priori -- all such 
information must be inferred by inspection of the VSH transforms.  This 
distinguishes the method from approaches to convective turbulence such as 
Reynolds-averaging of the hydrodynamic equations \citep{Viallet2013}, or the 
reduced-model approach based on ensembles of Lorenz convective rolls 
\citep{2011ApJ...741...33A}, where the dynamics are baked in to the analysis from 
the outset.  One consequence of the dynamical agnosticism of the VSH 
transform method is that a choice of phase distribution (such as our IID 
uniform choice) can fail to reproduce the coherent spatial structure (such 
as convective rolls) that appear in the simulation from which the VSH 
transform was extracted.

The spatial coherence corresponding to dynamical flow structures 
characteristic of convective turbulence necessarily requires some kind of 
phase coherence.  To reproduce it, the complex phases of the VSH transforms, 
as a function of $n,l,m$, should be modeled by some distribution other than 
IID uniform. Those phases are of course available from the VSH transform. It 
would be interesting -- and perfectly possible -- to study them, and 
possibly even attempt to model them empirically using some correlated 
stochastic model, generalizing the naive IID uniform treatment of \S\ref{Realization}.  
If the distributional parameters of such a model were found to be in some 
sense statistically stable, we would have a further grip on the statistical 
description of the flow, above and beyond the statistically-stable 
properties of the spectrum. This possibility is a promising avenue of future 
research in stellar convective turbulence.
Nevertheless, the velocity fields realized from VSH power spectra also obey to the requirements 
of the method on mass and momentum
conservation as well as the proper boundary conditions. In the future we plan to investigate
the sensitivity of CCSN explosion to phase restoration by
random drawings using different seeds.

It is interesting to contrast the VSH characterization of a turbulent flow 
with the Karhune-Loeve (K-L) decomposition \citep{1987QApMa..45..561S,1991JFM...222..231D}, an 
established method for extracting the statistical content of such a flow. 
The K-L decomposition, which extracts the principal eigenmodes of the 
empirical kernel constructed from many relizations of the flow, is capable 
of exhibiting the principal coherent structures that are present in the 
flow, and indeed of correlating velocity field structure with attending 
structure in other field variables.  The K-L decomposition therefore 
directly incoporates the phase-coherence modeling that would need to be 
added separately to the VSH decomposition, and therefore yields greater 
statistical information about the flow.  
%MANOS: COMMENTED THAT STATEMENT OUT
%That information, unfortunately 
%comes at considerable additional cost -- the empirical kernel must be 
%constructed from a large number of independent, well-aged realizations, 
%which are often unaffordable in the case of high-resolution 3D
%stellar simulations.  
In addition, for efficient decomposition, the kernel 
must be diagonalized by availing oneself of the problem symmetries.  For box 
flows, this means Fourier transforms.  In the spherical case of stars, the 
vector velocity component would necessarily have to be decomposed 
spherically, along the lines of a VSH decomposition. 
In summary, although the VSH decomposition method can itself be more
computationally expensive than the K-L decomposition 
since it requires full multi-dimensional simulations
and the calculation of a sufficient number of VSH modes it has it's advantages
with respect to the intended application to pre-SN stars with concentric convective shells.
%In summary, adopting a 
%VSH decomposition of a single simulated flow directly, as we do here, 
%accesses some of the statistical information of the flow at far lower cost 
%than would be incurred by a full K-L decomposition.

An additional limitation of the VSH method derives from the fact that the 
velocity field, even at very low Mach number, does not fully characterize a 
turbulent flow. Pressure and density perturbation fields of comparable 
complexity to the velocity field are necessarily also present. We do not 
discuss the density perturbation field in this paper, but it may in fact 
play an important role in the core-collapse dynamics resulting from Pre-SN 
turbulence. For example, \citet{2007ApJ...665..690M} show that the largest 
density perturbations can occur at convective boundaries, whereas the 
implementation of the VSH decomposition developed here necessarily employs 
"internal" boundary conditions, which protects the stratification but also 
eliminates the  velocity fluctuations associated with density fluctuations 
at these boundaries. The effects of those fluctuations are therefore 
excluded from explosions that start from VSH realizations.

It is worth noting, in any event, that from the point of view of preparing 
CCSN progenitor models of greater realism than radial inflow models, the 
IID-uniform realizations may be perfectly adequate.  There is no obvious 
specific reason why the difference between a spatially-coherent flow and a 
spatially incoherent one should be more important to the explosion 
properties of a CCSN than the difference between either of those and a 
radial flow.

We find that the slope of the VSH power spectrum of the solenoidal
modes in $l$ does not significantly change over the time-scales
we simulated with hints of power moving to different values of $l$
over time in both 2D and 3D.  This illustrates an advantage of the VSH spectral
decomposition: it furnishes a stable statistical description of the
turbulent convection in a star.

We also find that most of the power
goes to smaller radial and larger angular scales in 3D than in 2D. 
We determine the characteristic shape of the convective elements
to be more asymmetric in 3D than in 2D. More specifically the convective
eddies seem to be comparably elongated in the radial and tangential directions in 2D, while they
seem to be ``flatter'' and more elongated in the tangential directions in
3D. 
We stress, however, that these results should not be over-interpreted,
since the octant boundary conditions used in the 3D simulation result in
axis-aligned artifacts that probably affect the spectral content of the flow, 
and since we simulated at most one convective
overturn time-scale. We offer this analysis more as an illustration of the
power of the VSH method, and of the kind of general observations about the
nature of stratified convective flows that the method permits.

In the future we plan to use realized 3D convective fields obtained
by the VSH decomposition method 
to initialize core-collapse simulations and investigate
the effect of including more realistic initial conditions to the
core-collapse problem. We also plan to use the method to analyze
vigorous convective Si burning in the hours prior to core-collapse
and study its evolution with time with the aim of understanding where
most of the convective energy goes prior to collapse, and of
investigating the convective shell properties of a variety of SN
progenitors, including rotating and magnetized stars.

\acknowledgments

We would like to thank David Arnett, Donald Q. Lamb, Klaus Weide,
J. Craig Wheeler and the anonymous referee 
for useful discussions and comments.  EC would like
to thank the Enrico Fermi Institute for its support via the Enrico
Fermi Fellowship.  SMC is supported by NASA through Hubble Fellowship
grant No. 51286.01 awarded by the Space Telescope Science Institute,
which is operated by the Association of Universities for Research in
Astronomy, Inc., for NASA, under contract NAS 5-26555.  This work was
supported in part by the National Science Foundation under grant
AST-0909132.  The software used in this work was in part developed by
the DOE NNSA-ASC OASCR Flash Center at the University of Chicago.
This research used computational resources at ALCF at ANL, which is
supported by the Office of Science of the US Department of Energy
under Contract No. DE-AC02-06CH11357.  The authors acknowledge the
Texas Advanced Computing Center (TACC) at The University of Texas at
Austin for providing high-performance computing, visualization, and
data storage resources that have contributed to the research results
reported within this paper.

\appendix

\section{DECOMPOSITION OF A GENERAL VECTOR FIELD INTO VSH MODES}
\label{sec:appendix}

\subsection{Orthonormality and completeness.}

The vector fields $\mathbf{A}$, $\mathbf{B},$ $\mathbf{C}$ are
defined in terms of a generic Helmholtz equation solution $\phi_{k}$.
In order to furnish a full set of vector modes, it is necessary
to specify a full Sturm-Liouville family of such solutions.

Accordingly, we adopt the orthonormal family of Helmholtz equation
solutions
\begin{equation}
\phi_{nlm}(\mathbf{x})=g_{l}^{(b)}\left(k_{ln}^{(b)}r\right)Y_{lm}(\mathbf{n}),\label{eq:ScalarModes}
\end{equation}
where $Y_{lm}(\mathbf{n})$ is a spherical harmonic function, and
$g_{l}^{(b)}(k_{ln}^{(b)}r)$ is a solution of the radially-separated
Helmholtz equation,
\begin{equation}
\frac{1}{r^{2}}\frac{d}{dr}\left(r^{2}\frac{dg_{l}^{(b)}(k_{ln}^{(b)}r)}{dr}\right)+\left(k_{ln}^{(b)\,2}-\frac{l(l+1)}{r^{2}}\right)g_{l}(k_{ln}^{(b)}r)=0.\label{eq:Radial_Helmholtz}
\end{equation}

The superscript $b=D,N$ (``Dirichlet'' and ``Neumann'') on the
function $g_{l}^{(b)}$ and on the eigenvalue $k_{ln}^{(b)}$ refers
to alternative sets of boundary conditions:
\begin{equation}
g_{l}^{(D)}\left(k_{ln}^{(D)}R_{1}\right)=g_{l}^{(D)}\left(k_{ln}^{(D)}R_{2}\right)=0,\label{eq:BC_Type_1}
\end{equation}
and

\begin{equation}
\frac{dg_{l}^{(N)}}{dr}\left(k_{ln}^{(N)}r\right)\Biggl|_{r=R_{1}}=\frac{dg_{l}^{(N)}}{dr}\left(k_{ln}^{(N)}r\right)\Biggl|_{r=R_{2}}=0.\label{eq:BC_Type_2}
\end{equation}

The discrete index $n=0,1,2,\ldots$ on $k_{ln}^{(b)}$ arises because
the boundary conditions can only be satisfied for $k$ values in a
discrete set. It is necessary to consider both sets of boundary conditions
--- Dirichlet and Neumann --- because, as we will see, both types
are required to create vector modes that satisfy the boundary condition
of Equation~(\ref{eq:NoNetMomentum}).

The function $g_{l}^{(b)}(x)$ is necessarily a linear combination
of spherical Bessel functions of the first and second kind, $j_{l}(x)$
and $n_{l}(x)$. The precise linear combination is dictated (up to
a normalization) by the boundary conditions at $R_{1}$ and $R_{2}$.
The overall normalization is chosen so that
\begin{equation}
\int_{R_{1}}^{R_{2}}r^{2}dr\, g_{l}(k_{ln}r)g_{l}(k_{n^{\prime}}r)=\delta_{nn^{\prime}}.\label{eq:Radial_Functions_1}
\end{equation}
The specific form of these radial functions is exhibited in the next
section.

In terms of these modes, we define vector spherical modes patterned
after the fields $\mathbf{A}$, $\mathbf{B},$ $\mathbf{C}$ of Equations (\ref{eq:IrrotFun}--\ref{eq:SolenoidFun2}). The
irrotational mode $\mathbf{A}_{lnm}^{(b)}$ is 
\begin{eqnarray}
\mathbf{A}_{lnm}^{(b)}(\mathbf{x}) & \equiv & \lambda_{lnm}^{(b)}\nabla\left[g_{l}^{(b)}\left(k_{ln}^{(b)}r\right)Y_{lm}(\mathbf{n})\right]\nonumber \\
 & = & \lambda_{lnm}^{(b)}\left[\frac{dg_{l}^{(b)}\left(k_{ln}^{(b)}r\right)}{dr}\frac{\mathbf{x}}{r}Y_{lm}(\mathbf{n})+g_{l}^{(b)}\left(k_{ln}^{(b)}r\right)\nabla Y_{lm}(\mathbf{n})\right],\label{eq:A_Mode}
\end{eqnarray}
where $\lambda_{lnm}^{(b)}$ is a normalization constant to be determined.
It is noteworthy and useful in what follows that the vector $\nabla Y_{lm}(\mathbf{n})$
appearing in Equation~(\ref{eq:A_Mode}) is perpendicular to the radial
direction, $\mathbf{x}\cdot\nabla Y_{lm}(\mathbf{n})=0$.

On careful examination, we can see that with the boundary conditions
on $g_{l}^{(D)}$ and $g_{l}^{(N)}$ given in Equations~(\ref{eq:BC_Type_1}), (\ref{eq:BC_Type_2}), 
$\mathbf{A}_{lnm}^{(D)}$ is purely radial
at the boundaries, whereas $\mathbf{A}_{lnm}^{(N)}$ is purely tangential
at the boundaries. By the physical boundary conditions on momenta (Equation~\ref{eq:NoNetMomentum}) 
it follows that only the $\mathbf{A}_{lnm}^{(N)}$
modes may be legitimately employed to represent stratified velocity
fields. For whole-star ($R_{1}\rightarrow0$) fields, on the other
hand, both $\mathbf{A}$ modes are available, and the appropriate
mode should be selected depending on whether the velocity boundary
condition at $R_{2}$ represents outflow or a tangential motion.

The first solenoidal mode $\mathbf{B}_{lnm}^{(b)}$ is
\begin{eqnarray}
\mathbf{B}_{lnm}^{(b)}(\mathbf{x}) & \equiv & \eta_{lnm}^{(b)}\nabla\times\left[\mathbf{x}\, g_{l}^{(b)}\left(k_{ln}^{(b)}r\right)Y_{lm}(\mathbf{n})\right]\nonumber \\
 & = & -\eta_{lnm}^{(b)}\, g_{l}^{(b)}\left(k_{ln}^{(b)}r\right)\,\mathbf{x}\times\nabla Y_{lm}(\mathbf{n}).\label{eq:B_Mode}
\end{eqnarray}
Again, $\eta_{lmn}^{(b)}$ is a normalization constant. Both the Dirichlet
and Neumann scalar functions are available, since manifestly neither
leads to radial motion at the boundaries (or anywhere else, for that
matter). Only one of the two sets should be selected, however. Recall
that this mode represents a single degree of freedom (i.e. one scalar
function) satisfying definite boundary conditions at $R_{1}$ and
$R_{2}$. Either mode set has the coverage to represent such a function.
Given that the Dirichlet and Neumann modes aren't even mutually orthogonal
(see below), representing a vector field using both modes would be
problematic, and certainly non-unique. The Dirichlet mode leads to
velocity fields that are zero at the boundaries, so that if non-zero
tangential motion at the boundary must be modeled, the Neumann modes
are required. For this reason we will use the Neumann modes, setting\textbf{
$\mathbf{B}_{lnm}=\mathbf{B}_{lnm}^{(N)}$} by default.

The second solenoidal mode $\mathbf{C}_{lnm}^{(b)}$ is
\begin{eqnarray}
\mathbf{C}_{lnm}^{(b)} & \equiv & \chi_{lnm}^{(b)}\nabla\times\nabla\times\left[\mathbf{x}\, g_{l}^{(b)}\left(k_{ln}^{(b)}r\right)Y_{lm}(\mathbf{n})\right]\nonumber \\
 & = & \chi_{lnm}^{(b)}\nabla\times\left[-g_{l}^{(b)}\left(k_{ln}^{(b)}r\right)\,\mathbf{x}\times\nabla Y_{lm}(\mathbf{n})\right].\label{eq:C_Mode}
\end{eqnarray}
This can be expressed in terms of the elementary vectors $\mathbf{x}$
and $\nabla Y_{lm}$, using the identity $\nabla\times\left[\mathbf{a}\times\mathbf{b}\right]=\left(\mathbf{b}\cdot\nabla\right)\mathbf{a}-\left(\mathbf{a}\cdot\nabla\right)\mathbf{b}+\mathbf{a}\left(\nabla\cdot\mathbf{b}\right)-\mathbf{b}\left(\nabla\cdot\mathbf{a}\right)$:
\begin{eqnarray}
\mathbf{C}_{lnm}^{(b)}(\mathbf{x}) & = & -\chi_{lnm}^{(b)}\Biggl\{\nabla Y_{lm}(\mathbf{n})\cdot\left[\frac{dg_{l}^{(b)}\left(k_{ln}^{(b)}r\right)}{dr}\,\frac{\mathbf{xx}}{r}+g_{l}^{(b)}\left(k_{ln}^{(b)}r\right)\,\mathbf{1}\right]\nonumber \\
 &  & -g_{l}^{(b)}\left(k_{ln}^{(b)}r\right)\,\left[\nabla\left(\mathbf{x}\cdot\nabla Y_{lm}(\mathbf{n})\right)-\nabla Y_{lm}(\mathbf{n})\right]-g_{l}^{(b)}\left(k_{ln}^{(b)}r\right)\mathbf{x}\frac{l(l+1)}{r^{2}}Y_{lm}(\mathbf{n})\nonumber \\
 &  & -\nabla Y_{lm}(\mathbf{n})\left[r\frac{dg_{l}^{(b)}\left(k_{ln}^{(b)}r\right)}{dr}+3g_{l}^{(b)}\left(k_{ln}^{(b)}r\right)\right]\Biggr\},\label{eq:C_Mode_Alt1}
\end{eqnarray}
where we've used the identity $\nabla r=\mathbf{x}/r$, the commutator
$(\mathbf{x}\cdot\nabla)\nabla-\nabla(\mathbf{x}\cdot\nabla)=-\nabla$,
and the spherical Laplacian $\nabla^{2}Y_{lm}(\mathbf{n})=-\frac{l(l+1)}{r^{2}}Y_{lm}(\mathbf{n})$.
Using the fact that $\mathbf{x}\cdot\nabla Y_{lm}(\mathbf{n})=0$,
we obtain
\begin{eqnarray}
\mathbf{C}_{lnm}^{(b)}(\mathbf{x}) & = & \chi_{lnm}^{(b)}\left\{ \frac{l(l+1)}{r^{2}}\, Y_{lm}(\mathbf{n})\, g_{l}^{(b)}\left(k_{ln}^{(b)}r\right)\,\mathbf{x}+\left(r\frac{dg_{l}^{(b)}\left(k_{ln}^{(b)}r\right)}{dr}+g_{l}^{(b)}\left(k_{ln}^{(b)}r\right)\right)\nabla Y_{lm}(\mathbf{n})\right\} \nonumber \\
 & = & \chi_{lnm}^{(b)}\left\{ \frac{l(l+1)}{r^{2}}\, Y_{lm}(\mathbf{n})\, g_{l}^{(b)}\left(k_{ln}^{(b)}r\right)\,\mathbf{x}+\frac{d}{dr}\left[rg_{l}^{(b)}\left(k_{ln}^{(b)}r\right)\right]\nabla Y_{lm}(\mathbf{n})\right\} .\label{eq:C_Mode_Alt2}
\end{eqnarray}
It is clear from this expression that $\mathbf{}$$\mathbf{C}^{(N)}$
has a non-zero radial component at the boundaries, while $\mathbf{C}^{(D)}$
is purely tangential at the boundaries. We therefore choose the Dirichlet
modes $\mathbf{C}_{lnm}^{(D)}$ as the mode set appropriate to the
boundary condition of Equation~(\ref{eq:NoNetMomentum}). As we've now selected
a scalar boundary condition for each mode set, we henceforth we drop
the $N$ or $D$ subscripts from the mode vectors for convenience.
From here on, $\mathbf{A}_{lmn}=\mathbf{A}_{lmn}^{(N)}$, $\mathbf{B}_{lmn}=\mathbf{B}_{lmn}^{(N)}$,
and $\mathbf{C}_{lmn}=\mathbf{C}_{lmn}^{(D)}$. 
The expressions for $\mathbf{A}_{lnm}$, $\mathbf{B}_{lnm}$, $\mathbf{C}_{lnm}$
contained in Equations~(\ref{eq:A_Mode}), (\ref{eq:B_Mode}) and (\ref{eq:C_Mode_Alt2})
correspond to those in Equations 13.3.67-69 of \citet{1953mtp..book.....M}.

It can be shown that all modes satisfy orthogonality relations. The $\mathbf{A}_{lnm}$
are mutually orthogonal:
\begin{equation}
\int_{\Sigma}d^{3}\mathbf{x}\,\mathbf{A}_{lnm}(\mathbf{x})\cdot\mathbf{A}_{l^{\prime}n^{\prime}m^{\prime}}^{*}(\mathbf{x})  =  
\lambda_{lnm}^{2}k_{ln}^{(N)\,2}\delta_{nn^{\prime}}\delta_{ll^{\prime}}\delta_{mm^{\prime}},\label{eq:A_Mode_Orthogonality}
\end{equation}
where we've used
boundary condition on $g_{l^{\prime}}^{(N)}(k_{l^{\prime}n^{\prime}}^{(N)}r)=0$,
as well as the fact that $\mathbf{n}\cdot\nabla Y_{lm}=0$. To confer
unit norm upon the $\mathbf{A}_{lnm}$ modes we thus choose
\begin{equation}
\lambda_{lnm}=k_{ln}^{(N)\,-1}.\label{eq:A_Mode_Normalization}
\end{equation}

The $\mathbf{B}_{lnm}$ are also mutually orthogonal:
\begin{equation}
\int_{\Sigma}d^{3}\mathbf{x}\,\mathbf{B}_{lnm}(\mathbf{x})\cdot\mathbf{B}_{l^{\prime}n^{\prime}m^{\prime}}^{*}(\mathbf{x}) =
\eta_{lnm}^{2}\, l(l+1)\,\delta_{nn^{\prime}}\delta_{ll^{\prime}}\delta_{mm^{\prime}},\label{eq:B_Mode_Orthogonality}
\end{equation}
It follows that $\mathbf{B}_{lnm}$ has unit norm if 
\begin{equation}
\eta_{lnm}=\left[l(l+1)\right]^{-1/2}.\label{eq:B_Mode_Normalization}
\end{equation}
It is straightforward to show that this orthonormalization would be unchanged
had we selected Dirichlet modes instead of Neumann modes for $\mathbf{B}_{lnm}$. 

Finally, the $\mathbf{C}_{lnm}$ are mutually orthogonal as well:
\begin{equation}
\int_{\Sigma}d^{3}\mathbf{x}\,\mathbf{C}_{lnm}(\mathbf{x})\cdot\mathbf{C}_{l^{\prime}n^{\prime}m^{\prime}}^{*}(\mathbf{x}) =
\chi_{lnm}^{2}\, l(l+1)\, k_{ln}^{(D)\,2}\delta_{nn^{\prime}}\delta_{ll^{\prime}}\delta_{mm^{\prime}}\label{eq:C_Mode_Orthogonality}
\end{equation}
$\mathbf{C}_{lnm}$ has unit norm if
\begin{equation}
\chi_{lnm}=k_{ln}^{(D)\,-1}\left[l(l+1)\right]^{-1/2}.\label{eq:C_Mode_Normalization}
\end{equation}

It can also be shown that each set is orthogonal to the other two:
\begin{equation}
\int_{\Sigma}d^{3}\mathbf{x}\,\mathbf{A}_{lnm}(\mathbf{x})\cdot\mathbf{B}_{l^{\prime}n^{\prime}m^{\prime}}^{*}(\mathbf{x}) =
\int_{\Sigma}d^{3}\mathbf{x}\,\mathbf{A}_{lnm}(\mathbf{x})\cdot\mathbf{C}_{l^{\prime}n^{\prime}m^{\prime}}^{*}(\mathbf{x}) =
\int_{\Sigma}d^{3}\mathbf{x}\,\mathbf{B}_{lnm}(\mathbf{x})\cdot\mathbf{C}_{l^{\prime}n^{\prime}m^{\prime}}^{*}(\mathbf{x}) = 0.
\label{eq:Mutual_Orthogonality}
\end{equation}
Thus the three sets of modes are all mutually orthogonal, and each
of the $\mathbf{A}_{lnm}$, $\mathbf{B}_{lnm}$, $\mathbf{C}_{lnm}$
represents a set comprising mutually orthonormal modes.

Completeness has not been established here. However, it is plausible
to assume it. An arbitrary vector field corresponds in a sense to
three scalar functions, each of which may be completely decomposed
by sets of scalar functions such as the $\phi_{lnm}(\mathbf{x})$.
Since we have deployed a set of such scalar functions in each of the
three vector mode sets, the degree-of-freedom count is unvaried, and
we should expect to be able to match arbitrary vector functions with
these modes.

\subsection{Boundary conditions: The Dirichlet and Neumann radial modes.}

The Dirichlet radial mode functions $g_{l}^{(D)}(k_{n}r)$ satisfy
the radially-separated Helmholtz equation
\begin{equation}
\frac{1}{r^{2}}\frac{d}{dr}\left(r^{2}\frac{dg_{l}^{(D)}(kr)}{dr}\right)+\left(k^{2}-\frac{l(l+1)}{r^{2}}\right)g_{l}^{(D)}(kr)=0,\label{eq:RadialHelmholtz2}
\end{equation}
subject to the boundary conditions (B.C.)
\begin{equation}
g_{l}^{(D)}(k_{ln}^{(D)}R_{1})=g_{l}^{(D)}(k_{ln}^{(D)}R_{2})=0.\label{eq:BC2}
\end{equation}
As such, they are necessarily linear combinations of spherical Bessel
functions of the first and second kind, $j_{l}(k_{ln}^{(D)}r)$ and
$n_{l}(k_{ln}^{(D)}r)$. It is straightforward to write down a combination
that satisfies the B.C. at $R_{1}$:
\begin{eqnarray}
g_{l}^{(D)}(kr) & = & \mu_{ln}\,\left[n_{l}(kR_{1})j_{l}(kr)-j_{l}(kR_{1})n_{l}(kr)\right]\nonumber \\
 & = & \mu_{ln}\, T_{ll}(kR_{1},kr),\label{eq:g_1}
\end{eqnarray}
where $\mu_{ln}$ is a normalization constant, and where for
convenience we've introduced the notation
\begin{equation}
T_{l_{1}l_{2}}(x_{1},x_{2})\equiv n_{l_{1}}(x_{1})j_{l_{2}}(x_{2})-j_{l_{1}}(x_{1})n_{l_{2}}(x_{2}).\label{eq:T_Function}
\end{equation}

The $k_{ln}^{(D)}$ are the discrete, infinite set of roots of the
equation
\begin{equation}
Q_{l}(k)\equiv T_{l\, l}(kR_{1},kR_{2})=0,\label{eq:Root_Eqn}
\end{equation}
obtained by imposing the boundary condition at $R_{2}$ on the solution
of Equation~(\ref{eq:g_1}).

Using the properties of the spherical Bessel functions and the
recursion relation (\citet{1995mmp..book.....A}):
\begin{equation}
z_{l}^{\prime}(x)+\frac{l+1}{x}z_{l}(x)=z_{l-1}(x),\label{eq:BesselRecursion}
\end{equation}
where $z_{l}(x)$ stands for either $j_{l}(x)$ or $n_{l}(x)$,
it can be shown that the radial function normalization constant $\mu_{ln}$ is given by
\begin{equation}
\mu_{ln}=\left\{ \frac{R_{2}^{\,3}}{2}\left[T_{l\, l-1}(k_{ln}^{(D)}R_{1},k_{ln}^{(D)}R_{2})\right]^{2}-\frac{R_{1}^{\,3}}{2}\left[T_{l\, l-1}(k_{ln}^{(D)}R_{1},k_{ln}^{(D)}R_{1})\right]^{2}\right\} ^{-1/2}.\label{eq:Radial_Normalization2}
\end{equation}

The Neumann mode functions $g_{l}^{(N)}(kr)$ also satisfy Equation~(\ref{eq:RadialHelmholtz2}),
but with boundary conditions
\begin{equation}
\frac{dg_{l}^{(N)}}{dr}(k_{ln}^{(N)}r)\Biggl|_{r=R_{1}}=\frac{dg_{l}^{(N)}}{dr}(k_{ln}^{(N)}r)\Biggl|_{r=R_{2}}=0.\label{eq:Neumann_BC}
\end{equation}
We may write down a linear combination of $j_{l}$ and
$n_{l}$ that satisfy the boundary condition at $r=R_{1}$:
\begin{equation}
g_{l}^{(N)}\left(kr\right) = \xi_{ln}\left\{ T_{l-1\, l}\left(kR_{1},kr\right)-\frac{l+1}{kR_{1}}T_{l\, l}\left(kR_{1},kr\right)\right\} ,\label{eq:g_Neumann}
\end{equation}
where the derivative terms were replaced using the recursion relation (Equation~\ref{eq:BesselRecursion}).

The $k_{ln}^{(N)}$ are therefore
the infinite set of discrete roots of the equation
\begin{equation}
k\xi_{ln}\left\{ \frac{\partial T_{l-1\, l}(kR_{1},kr)}{\partial(kr)}-\frac{l+1}{kR_{1}}\frac{\partial T_{l\, l}(kR_{1},kr)}{\partial(kr)}\right\} \Biggl|_{r=R_{2}} = 0.\label{eq:NeumanSolns}
\end{equation}
Using the relation of Equation~(\ref{eq:BesselRecursion}) this becomes
\begin{equation}
S_{l}(k)\equiv T_{l-1\, l-1}\left(kR_{1},kR_{2}\right)-\frac{l+1}{kR_{2}}T_{l-1\, l}\left(kR_{1},kR_{2}\right)-\frac{l+1}{kR_{1}}T_{l\, l-1}\left(kR_{1},kR_{2}\right)+\frac{(l+1)^{2}}{k^{2}R_{1}R_{2}}T_{l\, l}\left(kR_{1},kR_{2}\right)=0,\label{eq:NBC_Eqn2}
\end{equation}
and the normalization constant
\begin{equation}
\xi_{ln}=\left\{ \frac{1}{2}\left[r^{3}\left(1-\frac{l(l+1)}{k_{ln}^{(N)\,2}r^{2}}\right)\left(T_{l-1\, l}\left(kR_{1},kr\right)-\frac{l+1}{kR_{1}}T_{l\, l}\left(kR_{1},kr\right)\right)^{2}\right]_{R_{1}}^{R_{2}.}\right\} ^{-1/2}.\label{eq:NeumannNorm2}
\end{equation}

Equations~(\ref{eq:Root_Eqn}) and (\ref{eq:NBC_Eqn2}) are oscillatory and have countably infinitely many roots, $k_{ln}$, that interleave with extrema that, in turn
 are roots of the derivatives of these equations. Our code uses the Newton-Raphson method to determine all roots and extrema
in succession from $n =$~0 to $n = n_{\mathrm{max}}$ where $n_{\mathrm{max}}$ is given by Equation~\ref{eq:nmax} (\S\ref{vsh}).

\subsection{Dependence on vector field dimensionality and simulation domain.}

There are fundamental differences in calculating VHS using 2D versus 3D vector field data. 
In the 2D case there is no need to evaluate the $m \neq$~0 modes because there is no information in the $\phi$ direction 
and rotational symmetry can be assumed (all $d/d\phi$ terms are zero). 
Therefore only the $A_{nl0}$, $B_{nl0}$, and $C_{nl0}$ modes are calculated.
In such case the convective elements can be thought as having a toroidal shape in 3D. As a result the total number
of modes calculated is only $(n_{\mathrm{max}}+1)(l_{\mathrm{max}}+1)$.

In addition, one has to be careful about the 2D vector field data set used to calculate VHS in order for orthonormality
to be established. The 2D data set needs to include data representing the whole volume of the convective shell
and not just a quadrant portion (with polar angle $0<\theta<\pi/2$). Omission of half of the data will lead
to erroneous treatment of orthonormality for the odd numbered (in $l$) irrotational modes 
and the even numbered (also in $l$) solenoidal modes. 
Therefore, if only quadrant 2D simulation data are available a full reconstruction of the data assuming
symmetry by $\theta =$~$\pi/2$ is necessary prior to the calculation
of VSH. Similarly, in 3D octant simulations a full reconstruction of
all data (full sphere) needs to be done assuming symmetry relations
prior to the calculation of VSH power spectra.

%%%%REFERENCES%%%%%%%
%
%

{}                     
%\end{references}

\end{document}